\documentclass[lettersize,journal]{IEEEtran}
\usepackage{amsmath,amsfonts}
\usepackage{amssymb}
\usepackage{wasysym}
\usepackage{stix}
\usepackage{tikz}
\usepackage{algorithmic}
\usepackage{algorithm}
\usepackage{array}
\usepackage[caption=false,font=normalsize,labelfont=sf,textfont=sf]{subfig}
\usepackage{textcomp}
\usepackage{stfloats}
\usepackage{url}
\usepackage{verbatim}
\usepackage{graphicx}
\usepackage{cite}
\usepackage{booktabs}
\usepackage{hyperref}
\newcommand{\etal}{\textit{et al}.}
\newcommand{\tikzcircle}[2][red,fill=red]{\tikz[baseline=-0.5ex]\draw[#1,radius=#2] (0,0) circle ;}

\begin{document}

\title{FLLIC: Functionally Lossless Image Compression}

\author{Xi~Zhang,~\IEEEmembership{Member,~IEEE}, and
        Xiaolin~Wu,~\IEEEmembership{Fellow,~IEEE}
        % <-this % stops a space
% \thanks{This work was supported in part by the National Natural Science Foundation of China (NSFC) and Natural Sciences and Engineering Research Council of Canada (NSERC).}
% <-this % stops a space
\thanks{X.~Zhang is with the ANGEL Lab, Nanyang Technological University, Singapore (email: xi.zhang@ntu.edu.sg).}
% <-this % stops a space
\thanks{X.~Wu is with the School of Computing and Artificial Intelligence, Southwest Jiaotong University, Chengdu, China (email: xwu510@gmail.com).}
}

% The paper headers
\markboth{Journal of \LaTeX\ Class Files,~Vol.~14, No.~8, August~2021}%
{Shell \MakeLowercase{\textit{et al.}}: A Sample Article Using IEEEtran.cls for IEEE Journals}

% \IEEEpubid{0000--0000/00\$00.00~\copyright~2021 IEEE}
% Remember, if you use this you must call \IEEEpubidadjcol in the second
% column for its text to clear the IEEEpubid mark.

\maketitle

%%%%%%%%% ABSTRACT
\begin{abstract}
    Recently, DNN models for lossless image coding have surpassed their traditional counterparts in compression performance, reducing the previous lossless bit rate by about ten percent for natural color images. 
    But even with these advances, mathematically lossless image compression (MLLIC) ratios for natural images still fall short of the bandwidth and cost-effectiveness requirements of most practical imaging and vision systems at present and beyond.
    To overcome the performance barrier of MLLIC, we question the very necessity of MLLIC. Considering that all digital imaging sensors suffer from acquisition noises, why should we insist on mathematically lossless coding, i.e., wasting bits to preserve noises? 
    Instead, we propose a new paradigm of joint denoising and compression called functionally lossless image compression (FLLIC), which performs lossless compression of optimally denoised images (the optimality may be task-specific). 
    Although not literally lossless with respect to the noisy input, FLLIC aims to achieve the best possible reconstruction of the latent noise-free original image.
    Extensive experiments show that FLLIC achieves state-of-the-art performance in joint denoising and compression of noisy images and does so at a lower computational cost.
\end{abstract}


%%%%%%%%% BODY TEXT

\section{Introduction}
\IEEEPARstart{A}ccompanying the exciting progress of modern machine learning with deep neural networks (DNNs), many researchers have published a family of end-to-end optimized DNN image compression methods in recent years. Most of these methods are rate-distortion optimized for lossy compression
~\cite{Balle2017_End,Theis2017_Lossy,Agustsson2017_Soft,Balle2018_Vari,Minnen2018_Joint,Mentzer2018_Cond,Lee2019_Context,Chen2020_Learned,hific,agdl,
he2021checkerboard,yang2021slimmable,kim2022joint,he2022elic,lee2022selective,addl,zou2022devil,zhang2023_lvqac}.
By design, they cannot perform lossless or near-lossless image compression even with an unlimited bit budget. 
More recently, a number of research teams embark on developing DNN lossless image compression methods, aiming at minimum code length 
\cite{mentzer2019practical,mentzer2020learning,kingma2019bit,townsend2019hilloc,hoogeboom2019integer,ho2019compression,zhang2020ultra,zhang2021iflow,zhang2021ivpf,kang2022pilc,lc-fdnet,zhang2024BPS}.
These authors apply various deep neural networks, such as 
autoregressive models \cite{van2016pixel,salimans2017pixelcnn++}, 
variational auto-encoder (VAE) models \cite{kingma2013auto} and 
normalizing flow models \cite{kobyzev2020normalizing} ,
to learn the unknown probability distribution of given image data, and entropy encode the pixel values by arithmetic coding driven by the learned probability models.  
These DNN models for lossless image coding have beaten the best of the traditional lossless image codecs in compression performance, reducing the lossless bit rate by about ten percent on natural color images.

The importance and utility of lossless image compression lie in a wide range of applications in computer vision and image communications, involving many technical fields, such as medicine, remote sensing, precision engineering and scientific research.  Imaging in high spatial, spectral and temporal resolutions is instrumental to discoveries and innovations.  As achievable resolutions of modern imaging technologies steadily increase, users are inundated by the resulting astronomical amount of image and video data.  For example, pathology imaging scanners can easily produce 1GB or more data per specimen. For the sake of cost-effectiveness and system operability (e.g., real-time access via clouds to high-fidelity visual objects), acquired raw images and videos of high resolutions in multiple dimensions must be compressed.

Unlike in consumer applications (e.g., smartphones and social media), where users are mostly interested in the appearlingness of decompressed images and can be quite oblivious to small compression distortions at the signal level, many technical fields demand the highest possible fidelity of decompressed images.  In the latter case, the current gold standard is mathematically lossless image compression (MLLIC).  But even with the advances of recent DNN-based lossless image compression methods, mathematically lossless compression ratios for medical and remote sensing images are only around 2:1, short of the requirements of bandwidth and cost-effectiveness for most practical imaging and vision systems at present and in near future.

In order to break the bottleneck of MLLIC in compression performance, we question the very necessity of MLLIC in the first place.  In reality, almost all digital sensors, for the purpose of imaging or otherwise, inherently introduce acquisition noises. Therefore, mathematically lossless compression is a false proposition at the outset, as it is counterproductive to losslessly code the noisy image, why waste bits to preserve all noises?  In contrast to MLLIC (or literally lossless to be more precise), a more principled approach is lossless compression of optimally denoised images (the optimality may be task specific). We call this new paradigm of joint denoising and compression functionally lossless image compression (FLLIC).  Although not literally lossless with respect to the noisy input, FLLIC aims to achieve the best possible reconstruction of the latent noise-free original image.  Information theoretically speaking, denoising reduces the entropy of noisy images and hence increases the compressibility at the source.

\begin{figure*}[t]
    \centering
    \includegraphics[width=0.95\linewidth]{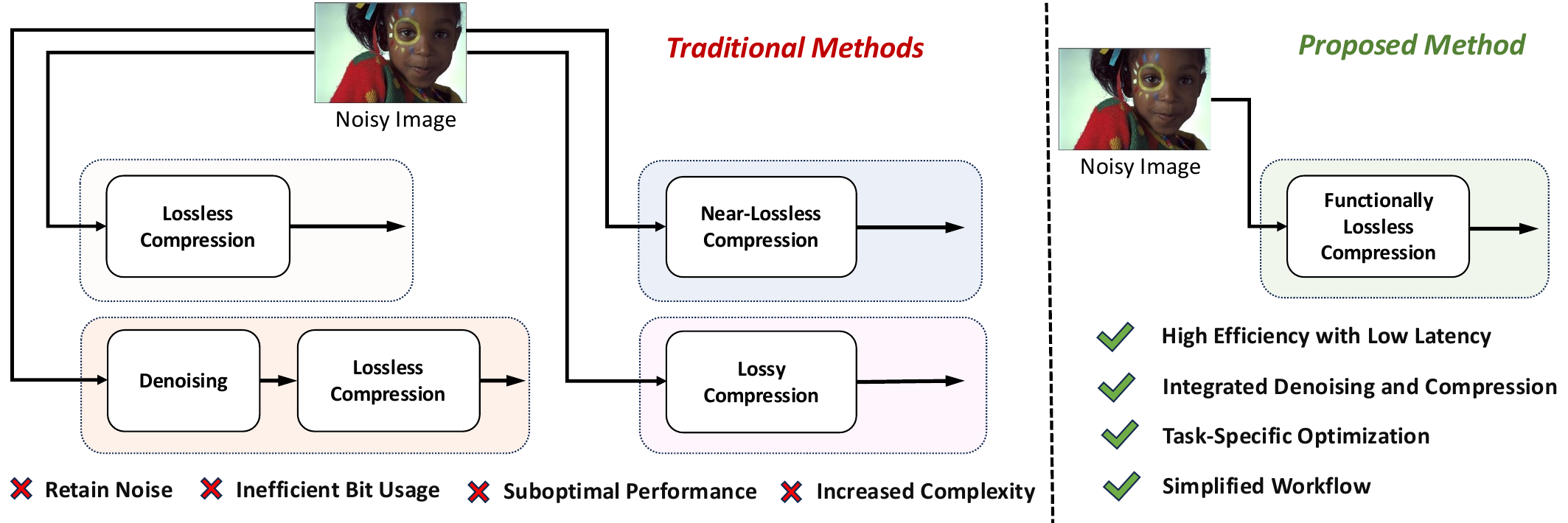}
    \caption{Comparison of traditional image compression methods with the proposed Functionally Lossless Image Compression (FLLIC) approach. Traditional methods either retain noise and waste bits (lossless compression), require complex processing steps (denoising followed by lossless compression), or result in suboptimal performance (near-lossless or lossy compression). In contrast, the proposed FLLIC method integrates denoising and compression into a unified, efficient, and low-latency framework, optimized for both task-specific performance and simplified workflows.}
    \label{fig:demo}
\end{figure*}

We provide a visual comparison between the traditional frameworks for noisy image compression and the proposed functionally lossless compression method in~\autoref{fig:demo}. In the current practice, a noisy image is either directly losslessly compressed or first denoised and then losslessly compressed. These two approaches are both sub-optimal in terms of rate-distortion metric. Direct lossless compression, by struggling to preserve noises, suffers in both aspects of fidelity and bit rate.  It is detrimental to the transmission and the subsequent machine vision tasks. The cascaded approach of denoising followed by lossless compression is complex and computationally expensive. In contrast, the proposed functionally lossless image compression (FLLIC) method jointly optimizes denoising and compression, and achieves higher coding efficiency and lower latency.

Our contributions are summarized as follows:
\begin{itemize}
    \item By exposing the limitations of current lossless image compression methods when dealing with noisy inputs, we introduce a new coding strategy of combining denoising and compression, called functionally lossless image compression (FLLIC).
    \item We propose and implement two different deep learning based solutions respectively for two scenarios: the latent clean image is available and unavailable in the training phase. 
    \item We provide a preliminary theoretical analysis of the relationship between the source entropy of clean image and its noisy counterpart, to support estimating the source entropy of clean image from its noisy observation.
    \item We propose an entropy-guided, content-adaptive quantization approach to achieve more efficient and accurate representations of latent clean image features.
    \item We conduct extensive experiments to show that the proposed functionally lossless compression method achieves state-of-the-art performance in joint denoising and compression of noisy images, outperforming the cascaded solution of denoising and compression, while requiring lower computational costs.
\end{itemize}

\section{Related Works}
\subsection{Lossy Image Compression}
Ballé \textit{et al.}~\cite{Balle2017_End} pioneered the first end-to-end convolutional neural network (CNN) model for image compression, marking a significant advancement in the field by integrating nonlinear transforms into the classical three-step signal compression framework: transform, quantization, and entropy coding. The key advantage of this CNN-based approach over traditional methods lies in replacing linear transforms with more powerful and flexible nonlinear transforms, effectively leveraging the representational capabilities of deep neural networks.

Building upon the foundational work of Ballé \textit{et al.}~\cite{Balle2017_End}, numerous end-to-end image compression methods have emerged, further improving rate-distortion (R-D) performance. These improvements stem from the design of more sophisticated nonlinear transforms and the development of more efficient context-based entropy models~\cite{rippel2017,agustsson2019,johnston2018,li2018}. In particular, adaptive context models for entropy estimation have gained significant attention~\cite{Mentzer2018_Cond,Balle2018_Vari,Minnen2018_Joint,Lee2019_Context}. For example, the CNN-based approaches by Minnen~\textit{et al.}~\cite{Minnen2018_Joint} and Lee \textit{et al.}~\cite{Lee2019_Context} achieved superior performance over the BPG codec in terms of PSNR.

In addition to improving entropy models, several studies have explored content-adaptive approaches that dynamically update encoder-side components during inference to further optimize compression efficiency~\cite{shin2022expanded,pan2022content,li2022content}. 
A substantial body of literature~\cite{Chen2020_Learned,lin2020spatial,davd,minnen2020channel,hific,
he2021checkerboard,yang2021slimmable,gao2021neural,kim2022joint,he2022elic,jiang2023mlic,jiang2023mlicpp,wang2023evc,zhang2024olvq} 
has been dedicated to enhancing R-D performance and coding efficiency through architectural innovations and refined quantization techniques.

More recently, transformer-based image compression methods~\cite{zhu2021transformer,lic-tcm} have emerged as strong alternatives to CNN-based approaches, demonstrating improved performance due to their global receptive field and ability to model long-range dependencies. Furthermore, diffusion model-based methods~\cite{cdc,careil2023towards} have begun to gain traction, showcasing their potential as next-generation solutions for image compression.
These developments collectively highlight the rapid progress in end-to-end learned image compression, driven by advancements in model architecture, entropy estimation, and adaptive inference techniques.

\subsection{Lossless and Near-lossless Image Compression}
Lossy and lossless image compression are two well-established topics that have been extensively studied for over two decades. However, there exists an intermediate area between lossy and lossless compression, known as near-lossless image compression (or \(\ell_\infty\)-constrained image compression)~\cite{ansari1998near}. The goal of near-lossless compression is to ensure that the absolute value of the compression error at each pixel is bounded by a user-specified error tolerance \(\tau\).

Several representative works have been proposed in this area:  
Zhou \etal~\cite{zhou2011,zhou2012} introduced a soft-decoding approach to reduce the \(\ell_2\) distortion of \(\ell_\infty\)-decoded images. Their method formulates soft decoding as an image restoration problem, leveraging the tight error bounds provided by \(\ell_\infty\)-constrained coding and employing a context-based modeler to handle quantization errors.  
Chuah \etal~\cite{chuah2013} replaced the traditional uniform scalar quantizer used in near-lossless image coding with a set of context-based, \(\ell_2\)-optimized quantizers. Their approach minimizes a weighted sum of \(\ell_2\) distortion and entropy while maintaining a strict \(\ell_\infty\) error bound.  
Li \etal~\cite{yuanman2014} proposed a sparsity-driven restoration technique to improve the coding performance of \(\ell_\infty\)-decoded images. Their method exploits the sparsity characteristics of the image for efficient restoration.  
Florea \etal~\cite{florea2016} focused on optimizing mesh codecs with respect to the \(\ell_\infty\) metric. They proposed novel data-dependent formulations for \(\ell_\infty\) distortion and incorporated \(\ell_\infty\)-based estimators into a state-of-the-art wavelet-based semi-regular mesh codec.  
Zhang \etal~\cite{zhang2020ultra} developed an asymmetric image compression system that features lightweight \(\ell_\infty\)-constrained predictive encoding combined with a computationally intensive, learning-based soft decoding process to achieve high-fidelity reconstruction.  Bai \etal~\cite{bai2021learning, bai2024deep} proposed a joint framework for lossy image compression and residual compression to achieve \(\ell_\infty\)-constrained near-lossless compression. 
Specifically, their approach first obtains a lossy reconstruction of the input image using lossy compression and then uniformly quantizes the residual to satisfy a tight \(\ell_\infty\) error bound.  
% These works collectively demonstrate the advancements in near-lossless image compression, focusing on balancing rate-distortion performance while strictly adhering to the \(\ell_\infty\) error constraint.

\subsection{Joint Image Compression and Denoising}
Image compression and image denoising 
%have been well-studied by both traditional methods and learning-based methods. 
have been thoroughly studied by researchers in both camps of traditional image processing and modern deep learning. 
However, the joint image compression and denoising task has been little explored. 
Very few papers addressed this topic. 
Testolina \etal \cite{testolina2021towards} investigated the integration of denoising convolutional layers in the decoder of a learning-based compression network.
Ranjbar \etal \cite{ranjbar2022joint} presented a learning-based image compression framework where image denoising and compression are performed jointly. The latent space of the image codec is organized in a scalable manner such that the clean image can be decoded from a subset of the latent space, while the noisy image is decoded from the full latent space at a higher rate. 
Cheng \etal \cite{cheng2022optimizing} proposed to optimize the image compression algorithm to be noise-aware as joint denoising and compression. The key is to transform the original noisy images to noise-free bits by eliminating the undesired noise during compression,
where the bits are later decompressed as clean images.
Huang \etal \cite{huang2023narv} proposed an efficient end-to-end image compression network, named Noise-Adaptive ResNet VAE (NARV), aiming to handle both clean and noisy input images of different noise levels in a single noise-adaptive compression network without adding nontrivial processing time.
Recent work~\cite{brummer2023importance} has explored the joint design of denoising and compression, demonstrating that integrating the two tasks yields superior rate--distortion trade-offs compared to cascaded pipelines. While sharing this high-level motivation, our method fundamentally differs in both objective and formulation. Specifically,~\cite{brummer2023importance} focuses on lossy compression trained with classical rate--distortion objectives, whereas our work establishes the new paradigm of \emph{Functionally Lossless Image Compression (FLLIC)}, which aims at lossless coding of optimally denoised content to achieve zero degradation for downstream tasks. 

Technically, FLLIC introduces entropy-guided, content-adaptive quantization driven by clean-entropy estimation and employs learned hyper-priors for precise entropy modeling. Moreover, we provide a theoretical analysis of the entropy gap between noisy and clean images to motivate our design---a perspective not covered in such prior works. 
% Finally, FLLIC reports bitrate under a strict lossless reconstruction criterion, whereas~\cite{brummer2023importance} evaluates standard lossy metrics such as PSNR and MS-SSIM.

% Although these works realized the significance of the joint image compression and denoising problem, they just combined image denoising with lossy compression task, with no regard to the lossless compression problem. To our best knowledge, we are the first to investigate the joint image denoising and lossless compression problem.

\section{Research Problems and Methodology}

\subsection{Problem Formulation}
Let \( I \) represent the latent clean image and \( I_n = I + N \) its noisy observation, where \( N \) accounts for sensor noise or other acquisition disturbances. The goal of functionally lossless image compression (FLLIC) is to estimate a reconstructed image \( \hat{I} \) from \( I_n \) while simultaneously minimizing both the distortion and the code length \( R(\hat{I}) \) required to encode \( \hat{I} \).

We address two settings: a supervised scenario, where the clean image \( I \) is accessible during training, and a weakly supervised scenario, where \( I \) is unavailable.

\medskip
\noindent\textbf{Scenario 1: Supervised Joint Compression and Denoising.}  
When the clean image \( I \) is available during training, the problem can be formulated as a supervised learning task. The objective is to jointly minimize the reconstruction distortion and the rate as follows:
\begin{equation}
\min_{\hat{I}} \| \hat{I} - I \| + \lambda R(\hat{I}),
\end{equation}
where \( \| \hat{I} - I \| \) measures the distortion between the reconstructed image \( \hat{I} \) and the clean image \( I \), \( R(\hat{I}) \) denotes the code length for representing \( \hat{I} \), and \( \lambda \) is a Lagrange multiplier that balances the trade-off between rate and distortion.

This formulation aligns with the principles of classical lossy image compression. However, it differs fundamentally as the input to the model is the noisy observation \( I_n \) instead of the clean image \( I \). The model must learn to denoise and compress simultaneously, leveraging the noisy data to approximate the latent clean image efficiently.

\medskip
\noindent\textbf{Scenario 2: Weakly Supervised Joint Compression and Denoising.}  
In practical scenarios, strictly noise-free images are often unavailable due to the inherent noise in most image acquisition processes. As a simplification, assume that the source entropy $H(I)$ of the clean image $I$ is known or can be approximated. We then use $H(I)$ as a form of weak supervision to guide the network toward a noise-free representation. In this setting, the objective becomes:
\begin{equation}
\min_{\hat{I}} \| \hat{I} - I_n \| + \lambda \| R(\hat{I}) - H(I) \|,
\end{equation}
where $\| \hat{I} - I_n \|$ ensures that the reconstructed image remains close to the noisy input, while enforcing $\| R(\hat{I}) - H(I) \|$ encourages the code length to approach the entropy of the clean image. By jointly minimizing these terms, the reconstructed image $\hat{I}$ is implicitly denoised and compressed, even without direct access to the clean image $I$ during training.

The architecture of the proposed entropy model guided by clean image entropy is illustrated in Fig.~2. Given a noisy input image \( I_n \), the system first extracts hierarchical features using a series of ResBlocks and Down-ResBlocks in the encoder. These blocks consist of convolutional layers that progressively downsample the input and extract spatially rich representations. A regression module estimates the entropy \( H(I) \) of the latent clean image \( I \), which serves as guidance for the entropy model. The Entropy Model Guided by Clean Image Entropy generates a Quantization Step Map, which determines spatial-channel-wise quantization steps for content-adaptive quantization. This adaptive quantization ensures that bits are allocated efficiently, focusing on regions with higher relevance to the image content.
The quantized features are then entropy-encoded (AE) for compression. During decoding, entropy-decoded (AD) features are passed through the decoder, which comprises Up-ResBlocks and ResBlocks. The Up-ResBlocks perform upsampling to restore the spatial resolution, while the ResBlocks further refine the reconstruction. The final output \( \hat{I} \) is a noise-reduced and compressed reconstruction of the original noisy input \( I_n \). The overall system achieves optimal joint denoising and compression by leveraging clean image entropy to guide the quantization process, thereby reducing entropy overhead and improving compression efficiency.

\begin{figure*}[t]
    \centering
    \includegraphics[width=0.98\textwidth]{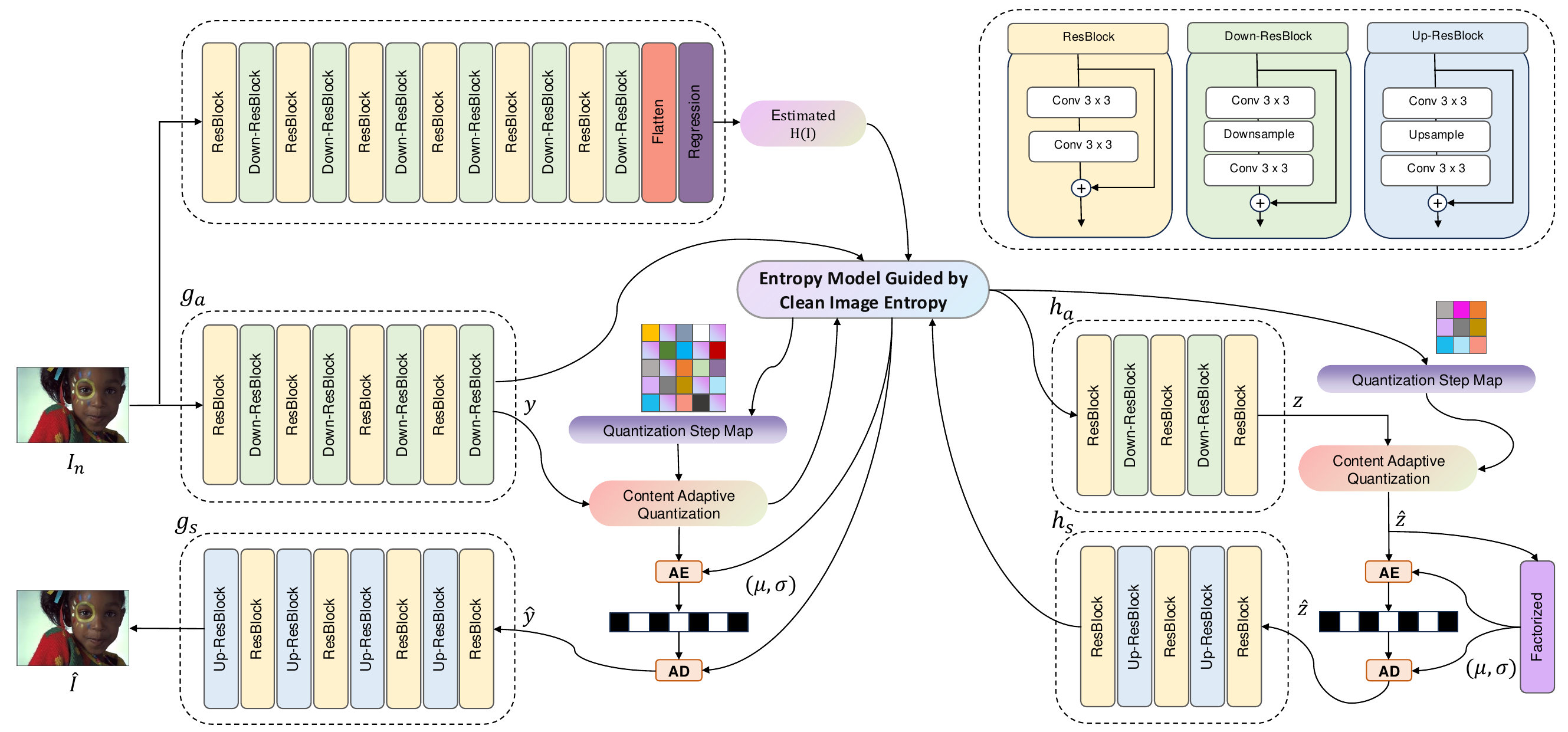}
    \caption{Architecture of the proposed FLLIC framework. The framework takes a noisy image as input and processes it through a series of ResBlocks and Down-ResBlocks to extract features and estimate the clean image entropy . The clean entropy estimation guides the entropy model to produce a Quantization Step Map for content-adaptive quantization. The quantized features are further entropy-encoded (AE) and entropy-decoded (AD) to reconstruct the output image . The decoder, composed of Up-ResBlocks and ResBlocks, refines the reconstruction. }
    \label{fig:framework}
\end{figure*}

\subsection{Theoretical Analysis of Clean Entropy Estimation}
In the context of FLLIC, we seek to minimize a code length or entropy measure associated with a latent clean image $I$ that is not directly accessible during training. Instead, we only have access to its noisy counterpart $I_n = I + N$, where $N$ represents additive noise. A key theoretical challenge is relating $H(I)$ to $H(I_n)$, or at least bounding $H(I)$ when only $I_n$ is observed. Doing so provides insight into why denoising prior to entropy coding can be fundamentally advantageous.

\textbf{Gaussian assumption.}
Let $X \in \mathbb{R}^n$ be a random vector representing the clean image, and $Y = X + N$ its noisy observation, where $N$ is an additive noise vector independent of $X$. We assume both $X$ and $N$ are zero-mean Gaussian vectors. Although real images are not strictly Gaussian, Gaussian modeling is a common and analytically tractable approximation that often provides valuable insight.
We denote the covariance matrices as $\Sigma_X = \mathbb{E}[XX^\top]$ and $\Sigma_N = \mathbb{E}[NN^\top]$. Without loss of generality, let $\Sigma_N = \sigma_N^2 I$, i.e., isotropic Gaussian noise with variance $\sigma_N^2$. Since $Y = X + N$, it follows that $\Sigma_Y = \Sigma_X + \Sigma_N$.

For a zero-mean Gaussian $X$ with covariance $\Sigma_X$, the differential entropy is
\begin{equation}
h(X) = \tfrac{1}{2}\log\bigl((2\pi e)^n \det(\Sigma_X)\bigr).
\end{equation}
Similarly,
\begin{equation}
h(Y) = \tfrac{1}{2}\log\bigl((2\pi e)^n \det(\Sigma_X + \Sigma_N)\bigr).
\end{equation}
Since $\Sigma_N$ is positive definite, $\det(\Sigma_X + \Sigma_N) > \det(\Sigma_X)$, implying:
\begin{equation}
h(Y) - h(X) = \tfrac{1}{2}\log\frac{\det(\Sigma_X + \Sigma_N)}{\det(\Sigma_X)} > 0.
\end{equation}
This inequality quantifies a fundamental cost of noise: observing a noisy image $Y$ instead of $X$ increases uncertainty and thus the entropy.

Let $\lambda_1 \ge \lambda_2 \ge \cdots \ge \lambda_n$ be the eigenvalues of $\Sigma_X$. Then
\begin{equation}
h(Y) - h(X) = \sum_{i=1}^n \tfrac{1}{2}\log\left(1 + \frac{\sigma_N^2}{\lambda_i}\right).
\label{eq:gauss_gap}
\end{equation}
This relationship links the entropy gap to the signal-to-noise ratio (SNR) characteristics of $X$. Eigenvalues $\lambda_i$ represent the variance of $X$ along principal components. Signal components with smaller $\lambda_i$ are more easily overwhelmed by noise, thus increasing $h(Y)$ relative to $h(X)$.

\textbf{Non-Gaussian extension.}
Now let \( X \) be an \emph{arbitrary} zero-mean vector with covariance 
\(\Sigma_X\), still independent of the Gaussian noise 
\( N \sim \mathcal{N}(0,\sigma_N^2 I) \).
Denote by \( g_Z \) the Gaussian with the same mean and covariance as a random vector \( Z \).
It holds that
\begin{equation}
  D(Z \parallel g_Z) = \frac{1}{2} \log\!\bigl[(2\pi e)^n \det(\Sigma_Z)\bigr] - h(Z).
\end{equation}
Using this for both \( X \) and \( Y \), we get
% \begin{equation}
%   h(Y) - h(X)
%   = \frac{1}{2} \log 
%   \frac{\det(\Sigma_X + \sigma_N^2 I)}{\det(\Sigma_X)}
%   + \bigl[\, D(X \parallel g_X) - D(Y \parallel g_Y) \bigr].
%   \label{eq:decomp}
% \end{equation}
\begin{equation}
\begin{split}
  h(Y) - h(X)
  &= \frac{1}{2} \log 
     \frac{\det(\Sigma_X + \sigma_N^2 I)}{\det(\Sigma_X)} \\[6pt]
  &\quad + \bigl[\, D(X \parallel g_X) - D(Y \parallel g_Y) \bigr].
\end{split}
\label{eq:decomp}
\end{equation}
Because adding independent Gaussian noise makes a distribution more Gaussian,
\[
  D(Y \parallel g_Y) \le D(X \parallel g_X),
\]
which implies the sharp lower bound
\begin{equation}
  h(Y) - h(X) 
  \;\ge\; 
  \frac{1}{2} \log 
  \frac{\det(\Sigma_X + \sigma_N^2 I)}{\det(\Sigma_X)}.
  \label{eq:nonGauss_lower}
\end{equation}
Equality holds if and only if \( X \) is Gaussian.

\textbf{Entropy–power inequality (EPI).}
Define the entropy power by
\[
  \mathcal{E}(Z) = \frac{1}{2\pi e} \exp\!\Bigl(\frac{2}{n} h(Z)\Bigr).
\]
Then the EPI states that for any independent \( X \) and \( N \),
\begin{equation}
  \mathcal{E}(X+N) \ge \mathcal{E}(X) + \mathcal{E}(N),
\end{equation}
which implies
\begin{equation}
  h(Y) - h(X)
  \;\ge\;
  \frac{n}{2} \log\!\Bigl( 1 + \frac{\mathcal{E}(N)}{\mathcal{E}(X)} \Bigr) > 0.
\end{equation}
For \( N \sim \mathcal{N}(0,\sigma_N^2 I) \), we have \( \mathcal{E}(N) = \sigma_N^2 \).
Since \( D(Y \parallel g_Y) \ge 0 \), the decomposition Eq.~\eqref{eq:decomp} implies
\begin{equation}
  h(Y) - h(X)
  \;\le\;
  \frac{1}{2} \log 
  \frac{\det(\Sigma_X + \sigma_N^2 I)}{\det(\Sigma_X)}
  + D(X \parallel g_X).
  \label{eq:nonGauss_upper}
\end{equation}
Equality holds if and only if \( X \) is Gaussian.

\textbf{Summary.}
Combining these facts, for arbitrary \( X \) we obtain a sharp sandwich:
% \begin{equation}
%   \boxed{
%     \frac{1}{2} \log 
%     \frac{\det(\Sigma_X + \sigma_N^2 I)}{\det(\Sigma_X)}
%     \;\le\;
%     h(Y) - h(X)
%     \;\le\;
%     \frac{1}{2} \log 
%     \frac{\det(\Sigma_X + \sigma_N^2 I)}{\det(\Sigma_X)}
%     + D(X \parallel g_X).
%   }
% \end{equation}
\begin{equation}
\begin{split}
  &\frac{1}{2} \log 
  \frac{\det(\Sigma_X + \sigma_N^2 I)}{\det(\Sigma_X)}
  \;\le\;
  h(Y) - h(X)\\
  &\le\;
  \frac{1}{2} \log 
  \frac{\det(\Sigma_X + \sigma_N^2 I)}{\det(\Sigma_X)}
  + D(X \parallel g_X).
\end{split}
\end{equation}

The gap equals exactly the Gaussian value Eq.~\eqref{eq:gauss_gap} 
if and only if \( X \) is Gaussian.
Building on this understanding, we can derive the following key insights and analyses:

\textbf{1. Coding inefficiency without denoising:}  
The term $h(Y) - h(X)$ represents the minimal entropy overhead due to noise. If one codes $Y$ directly without denoising, this overhead cannot be avoided. It sets a fundamental limit on the efficiency of compression when noise is present.

% \medskip
\textbf{2. Justification for the FLLIC paradigm:}  
By performing denoising to approximate $X$ from $Y$, we aim to reduce the effective entropy from $h(Y)$ down toward $h(X)$. Even imperfect denoising can lower the overall entropy, thereby improving compression performance and reducing code length. The larger the gap $h(Y)-h(X)$, the more benefit we gain by denoising.

% \medskip
\textbf{3. Connection to rate-distortion theory:}  
Classical rate-distortion theory posits that the minimal achievable rate for reconstructing a signal to within a certain distortion level depends on the signal’s distribution. As $Y$ is a noisy observation of $X$, removing noise can be viewed as reducing distortion relative to a noise-free reference. The entropy gap in Eq.~\eqref{eq:gauss_gap} suggests that approaching $X$ from $Y$ moves us closer to the fundamental lower bound on encoding cost.

% \medskip
\textbf{4. Small-noise approximation:}  
When the noise is small, which means $\sigma_N^2 \ll \lambda_n$, we have the approximation:
\begin{equation}
\sum_{i=1}^{n}\tfrac{1}{2}\log\left(1+\frac{\sigma_N^2}{\lambda_i}\right) \approx \sum_{i=1}^{n}\frac{\sigma_N^2}{2\lambda_i}.
\end{equation}
This linear approximation highlights that when noise is weak relative to the image variance, the entropy gap scales roughly as the sum of $\sigma_N^2/\lambda_i$, making denoising particularly effective in low-SNR regions of the image.

In practice, images are discrete, and we often consider the entropy $H(X)$ under fine quantization. If $X^\Delta$ is $X$ quantized with step size $\Delta$, then for sufficiently small $\Delta$:
\begin{equation}
\begin{aligned}
    & H(X^\Delta) \approx h(X) - n\log\Delta, \\
    & H(Y^\Delta) \approx h(Y) - n\log\Delta.
\end{aligned}
\end{equation}
Subtracting these gives $H(Y^\Delta)-H(X^\Delta) \approx h(Y)-h(X)$, showing that the differential-entropy-based analysis carries over to discrete-domain scenarios. Thus, the theoretical results remain relevant to practical coding, reinforcing the advantage of denoising before compression.

Our theoretical analysis provides a principled understanding of why FLLIC improves compression efficiency. The entropy gap $h(Y)-h(X)$, which quantifies how much noise inflates the source entropy, serves as a theoretical guide. By denoising, we effectively reduce this gap, approaching the fundamental entropy limit of the clean image, and thereby enabling more efficient and cost-effective compression.

\subsection{Practical Estimation of Clean Image Entropy}

Estimating the clean image entropy $H(I)$ in practice is non-trivial. We propose a deep neural network (DNN) to estimate $H(I)$ from the noisy observation $I_n$, an estimate of $H(I_n)$, and (optionally) the noise variance $\sigma_N^2$. This leverages both the statistical properties of the noisy image and auxiliary entropy information.

Given a clean image $I \in \mathbb{R}^{H \times W \times C}$ and additive Gaussian noise $N \sim \mathcal{N}(0, \sigma_N^2 I)$ with $I_n = I + N$, we first estimate:
\begin{equation}
H(I_n) = \hat{H}(I_n),
\end{equation}
where $\hat{H}(I_n)$ is obtained via mathematically lossless compression (MLLIC). Our DNN aims to approximate:
\begin{equation}
\hat{H}(I) = f_\phi(I_n, \hat{H}(I_n), \sigma_N),
\end{equation}
with parameters $\phi$.

The network consists of two modules. First, a feature extractor $g_\theta(\cdot)$ with parameters $\theta$ processes $I_n$ through cascaded residual (ResBlocks) and downsampling residual blocks (Down-ResBlocks) to obtain:
\begin{equation}
F(I_n) = g_\theta(I_n).
\end{equation}
ResBlocks and Down-ResBlocks follow standard definitions with residual connections and optional downsampling to capture hierarchical features and noise characteristics.

The output feature $F(I_n)$ is flattened into $F_v = \mathrm{vec}(F(I_n))$. We concatenate $F_v$, $\hat{H}(I_n)$, and $\sigma_N$ into:
\begin{equation}
Z = [F_v; \hat{H}(I_n); \sigma_N],
\end{equation}
which is passed to a regression module $r_\psi(\cdot)$ (an MLP with parameters $\psi$) to predict:
\begin{equation}
\hat{H}(I) = r_\psi(Z).
\end{equation}

We train the network by minimizing:
\begin{equation}
\min_{\theta,\psi} \mathbb{E}[(\hat{H}(I)-H(I))^2],
\end{equation}
using datasets of $(I, I_n, \sigma_N)$ with known or high-fidelity $H(I)$.

In summary, this framework combines learned features from noisy inputs with entropy priors to effectively estimate clean image entropy under practical conditions.

\subsection{Content-Adaptive Quantization and Probability Estimation}
\begin{figure}[t]
    \centering
    \includegraphics[width=0.95\linewidth]{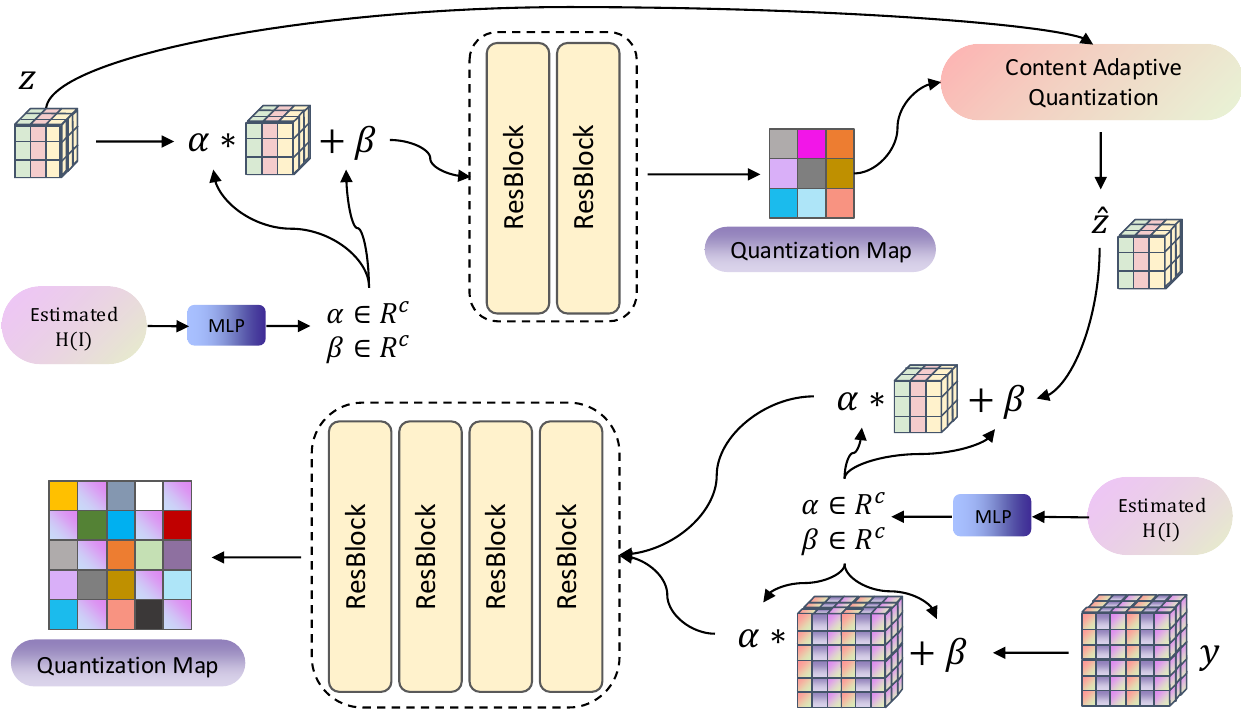}
    \caption{Illustration of the content-adaptive quantization process. Separate neural networks generate quantization step maps for the latent features $\mathbf{z}$ and hyperlatents $\mathbf{y}$, respectively.}
    \label{fig:content_adaptive_quant}
\end{figure}

The proposed framework integrates entropy-guided adaptive quantization (see~\autoref{fig:content_adaptive_quant}) and probability estimation to optimize joint denoising and compression. By incorporating clean image entropy $\hat{H}(I)$, latent features $\mathbf{y}$, and hyperlatents $\mathbf{z}$, the framework achieves precise control over quantization and accurate probability modeling, effectively balancing rate-distortion trade-offs.

\medskip
\textbf{Content-Adaptive Quantization.}  
To enable efficient representation of image features, we employ two distinct content-adaptive quantization step maps: one for the latent features $\mathbf{y}$ and another for the hyperlatents $\mathbf{z}$. These maps are dynamically computed to adapt to the statistical properties and complexity of the image content.

1. Quantization Map for Latent Features $\mathbf{y}$:  
   The quantization step map $\mathbf{Q}_\mathbf{y} \in \mathbb{R}^{H \times W \times C}$ for the latent features $\mathbf{y}$ is computed as:
   \(
   \mathbf{Q}_\mathbf{y} = f_\mathbf{Q_y}(\mathbf{y}, \hat{\mathbf{z}}, \hat{H}(I)),
   \)
   where $\mathbf{y} \in \mathbb{R}^{H \times W \times C}$ represents the latent features, $\hat{\mathbf{z}}$ denotes the decoded hyperlatents, and $\hat{H}(I)$ is the estimated clean image entropy. The function $f_\mathbf{Qy}(\cdot)$ is a neural network parameterized to produce $\mathbf{Q}_\mathbf{y}$ by analyzing the spatial, channel-wise, and entropy-based characteristics of $\mathbf{y}$. 

   The quantization process for $\mathbf{y}$ is defined as:
   \(
   \mathbf{y}_q = \mathrm{round}\left( \mathbf{y} / \mathbf{Q}_\mathbf{y} \right),
   \)
   where division and rounding are performed element-wise. Regions with higher complexity or structural importance are assigned finer quantization steps, ensuring accurate reconstruction, while smoother regions are quantized more coarsely to save bits.

2. Quantization Map for Hyperlatents $\mathbf{z}$: 
   The quantization step map $\mathbf{Q}_\mathbf{z} \in \mathbb{R}^{H' \times W' \times C'}$ for the hyperlatents $\mathbf{z}$ is derived as:
   \(
   \mathbf{Q}_\mathbf{z} = f_\mathbf{Q_z}(\mathbf{z}, \hat{H}(I)),
   \)
   where $\mathbf{z} \in \mathbb{R}^{H' \times W' \times C'}$ represents the hyperlatents, and $f_\mathbf{Qz}(\cdot)$ is a neural network conditioned on $\mathbf{z}$ and $\hat{H}(I)$. This design ensures that the quantization map adapts to the global statistical properties of $\mathbf{z}$ and the entropy of the clean image.

   The quantization process for $\mathbf{z}$ follows:
   \(
   \mathbf{z}_q = \mathrm{round}\left( \mathbf{z} / \mathbf{Q}_\mathbf{z} \right).
   \)
   This adaptive quantization efficiently encodes the hyperlatents while minimizing distortion.

\medskip
\textbf{Probability Distribution Estimation.}  
Efficient entropy coding is achieved by modeling the distributions of both latent features $\mathbf{y}$ and hyperlatents $\mathbf{z}$ with appropriate probability models.

1. Latent Feature Probability Estimation:
   The probability distribution parameters $(\mu_\mathbf{y}, \sigma_\mathbf{y})$ for $\mathbf{y}$ are estimated as:
   \begin{equation}
   (\mu_\mathbf{y}, \sigma_\mathbf{y}) = f_{\text{dist},\mathbf{y}}(\hat{\mathbf{z}}, \hat{H}(I)),
   \end{equation}
   where $f_{\text{dist},\mathbf{y}}(\cdot)$ is a neural network that uses the decoded hyperlatents $\hat{\mathbf{z}}$ and the estimated clean image entropy $\hat{H}(I)$ to predict the mean and variance of a Gaussian distribution:
   \begin{equation}
   p(\mathbf{y}_q \mid \hat{\mathbf{z}}, \hat{H}(I)) = \mathcal{N}(\mu_\mathbf{y}, \sigma_\mathbf{y}).
   \end{equation}

2. Hyperlatent Feature Probability Estimation:  
   For hyperlatents \( \mathbf{z} \), a factorized entropy model is employed:
   \begin{equation}
   p(\mathbf{z}_q) = \prod_{i} p(\mathbf{z}_q[i]),
   \end{equation}
   where \( p(\mathbf{z}_q[i]) \) represents independent univariate distributions. This simplification facilitates efficient entropy coding for hyperlatents.

\medskip
\textbf{Integrated Framework.}  
By integrating clean image entropy $\hat{H}(I)$, latent features $\mathbf{y}$, and hyperlatents $\mathbf{z}$, the proposed framework achieves an optimal balance between rate and distortion. The entropy-guided quantization maps $\mathbf{Q}_\mathbf{y}$ and $\mathbf{Q}_\mathbf{z}$ dynamically allocate bits to regions of higher structural importance, preserving critical image details with precision. Simultaneously, the probability estimation mechanism ensures accurate entropy coding, minimizing the bit rate required for high-fidelity reconstruction. This seamless interplay between adaptive quantization and robust probability modeling significantly enhances the efficiency and effectiveness of joint denoising and compression, reducing redundancy while focusing on essential image structures.

\section{Experiments}
In this section, we present the implementation details of the proposed FLLIC compression system. To systematically evaluate and analyze the performance of the FLLIC compression system, we conduct extensive experiments and compare our results with several state-of-the-art methods on quantitative metric and inference complexity.

\subsection{Experiment Setup}
In this part, we describe the experiment setup including the following four aspects: dataset preparation, training details, baselines and metrics. 

\textbf{Synthetic Datasets:}  
Following previous work on lossless image compression~\cite{lc-fdnet}, we employ the Flickr2K dataset~\cite{flicker2k}, which comprises 2,000 high-quality images, to train our proposed model. To thoroughly evaluate our approach under controlled noise conditions, we create synthetic noisy test sets by adding additive white Gaussian noise (AWGN) at four relatively small noise levels ($\sigma=1,2,3$) to two widely-used benchmark datasets: Kodak24~\cite{kodak24} and DIV2K validation dataset~\cite{div2k}. By varying the noise level in a fine-grained manner, we can systematically analyze the performance of our method in scenarios more reflective of practical imaging conditions than those typically used in purely denoising-oriented benchmarks.

\textbf{Real-World Dataset:}  
To further demonstrate the practical applicability and robustness of our FLLIC compression method, we also evaluate on the Smartphone Image Denoising Dataset (SIDD)~\cite{sidd}, which contains approximately 30,000 noisy images captured in real-world environments using five representative smartphone cameras. Each camera-image pair spans multiple scenes and lighting conditions, providing a realistic spectrum of noise characteristics not easily replicated by synthetic noise models. SIDD also includes "noise-free" reference images, allowing us to rigorously assess both the denoising fidelity and compression efficiency of the proposed approach in genuine, hardware-based noise scenarios. This evaluation ensures that our method not only excels in controlled synthetic settings but also translates effectively to challenging, real-world conditions.

\textbf{Training Details.} During training, we randomly extract patches of size \(256 \times 256\) from the images. The Adam optimizer~\cite{adam} is used with \(\beta_1 = 0.9\) and \(\beta_2 = 0.999\), and a batch size of 128. The initial learning rate is set to \(1 \times 10^{-4}\), and it decays by a factor of 0.5 every \(4 \times 10^{4}\) iterations, ultimately reaching \(1.25 \times 10^{-5}\). The model is trained with PyTorch on a NVIDIA RTX 4090 GPU, and convergence takes approximately two days.
Before training the FLLIC network, we first train the entropy estimation network using the DIV2K~\cite{div2k} dataset with the same training strategy described above. For the synthetic denoising dataset, we train a specific network for each noise level. In this case, an optimal Lagrange multiplier \(\lambda\) is selected for each noise level.

\textbf{Metrics.}
To evaluate the compression performance of the proposed FLLIC method, we use the rate-distortion metric, which considers both the bitrate (BPSP) and the reconstruction quality (PSNR). In the case of ideal lossless compression, the reconstructed image is identical to the original, resulting in zero distortion. However, in the FLLIC framework, the reconstructed image is compared to the latent noise-free image, rather than the noisy input image. This is because FLLIC aims to estimate the clean image from its noisy observation, so it is essential to evaluate the distortion in relation to the underlying noise-free image rather than the noisy input. Therefore, the rate-distortion metric used in FLLIC provides a more accurate reflection of the actual compression performance, as it quantifies both the compression efficiency (rate) and the quality of the denoising and compression process (distortion).

\textbf{Baselines.}
To thoroughly evaluate the performance of the proposed FLLIC approach, we compare it against four representative coding frameworks: (1) pure lossless compression, (2) a cascaded combination of denoising followed by lossless compression, (3) learned lossy compression, and (4) near-lossless compression. By considering a diverse set of baselines, we aim to more comprehensively assess the advantages of joint denoising and compression under different operational constraints.

\textbf{Baseline 1: Lossless Compression.}  
We use LC-FDNet~\cite{lc-fdnet} and ArIB-BPS~\cite{zhang2024BPS}, two state-of-the-art learning-based lossless image compression models as the baseline for pure lossless coding. LC-FDNet encodes images in a coarse-to-fine manner, separating low- and high-frequency components to achieve state-of-the-art compression performance on clean images. ArIB-BPS introduces Bit Plane Slicing (BPS) to enhance the autoregressive initial bits (ArIB) framework by splitting images in the bit plane dimension, with a dimension-tailored autoregressive model that efficiently captures dependencies across different dimensions.

\textbf{Baseline 2: Cascaded Denoising + Lossless Compression.}  
In this baseline, we explore three different cascaded approaches, where a denoising model is first applied to the noisy image before lossless compression. These combinations serve as natural extensions of the pure lossless baseline and leverage prior denoising techniques to potentially reduce the entropy of the input signal, thereby improving subsequent coding efficiency. The following three combinations are considered:

a). BM3D~\cite{bm3d} + FLIF~\cite{flif}: BM3D, a well-known traditional denoising algorithm, is paired with FLIF, the best traditional lossless image compression method, to first reduce noise and then compress the denoised image.

b). Restormer~\cite{restormer} + LC-FDNet~\cite{lc-fdnet}: Restormer, a transformer-based image restoration model, is used for denoising, followed by LC-FDNet for lossless compression.

c). CGNet~\cite{cgnet} + ArIB-BPS~\cite{zhang2024BPS}: CGNet, a state-of-the-art deep image denoising network, is combined with ArIB-BPS, a bit-plane slicing based lossless compression method, for denoising and compression.

% These cascaded methods are evaluated to understand the impact of different denoising and compression combinations on the overall performance.

\textbf{Baseline 3: Lossy Compression.}  
For comparison, we evaluate two state-of-the-art learned lossy compression frameworks that are inspired by end-to-end optimized neural image compressors (e.g., Minnen \emph{et al.}~\cite{Minnen2018_Joint}). These methods are designed to minimize a rate-distortion objective directly, making them powerful baselines for assessing the performance of our FLLIC approach. The two lossy compression methods considered are:

a). ELIC~\cite{he2022elic}: Efficient Learned Image Compression (ELIC) is a state-of-the-art framework that focuses on optimizing lossy compression by leveraging efficient entropy models and adaptive coding strategies.

b). MLIC~\cite{jiang2023mlic}: Multi-Reference Entropy Model for Learned Image Compression (MLIC) introduces a multi-reference entropy model to improve rate-distortion performance for lossy image compression.

% These lossy compression methods serve as strong baselines to demonstrate how FLLIC performs in comparison to advanced lossy codecs that do not specifically target noise removal.

\textbf{Baseline 4: Near-Lossless Compression.}  
To further evaluate the performance of the proposed FLLIC approach, we include a near-lossless compression framework as a point of comparison. This baseline allows for controlled, minimal distortion while still targeting efficient compression. For this, we employ the Deep Lossy Plus Residual (DLPR) Coding method~\cite{bai2021learning,bai2024deep}, which enables a trade-off between perfect fidelity and compression efficiency. The DLPR method is a state-of-the-art approach that balances maintaining near-lossless quality while optimizing coding performance, making it an ideal benchmark for comparison with FLLIC.

For a fair comparison, we fine-tune all baselines on the Flickr2K dataset. Pre-trained weights for each chosen method are used as initialization, and the entire fine-tuning process is conducted under conditions similar to those used for our FLLIC method. 
% This ensures that all baselines are well-adapted to the noisy or near-lossless input scenarios, allowing for a rigorous and meaningful evaluation.

\begin{figure*}[!t]
    \centering
    \begin{minipage}{0.329\linewidth}
        \includegraphics[width=\linewidth]{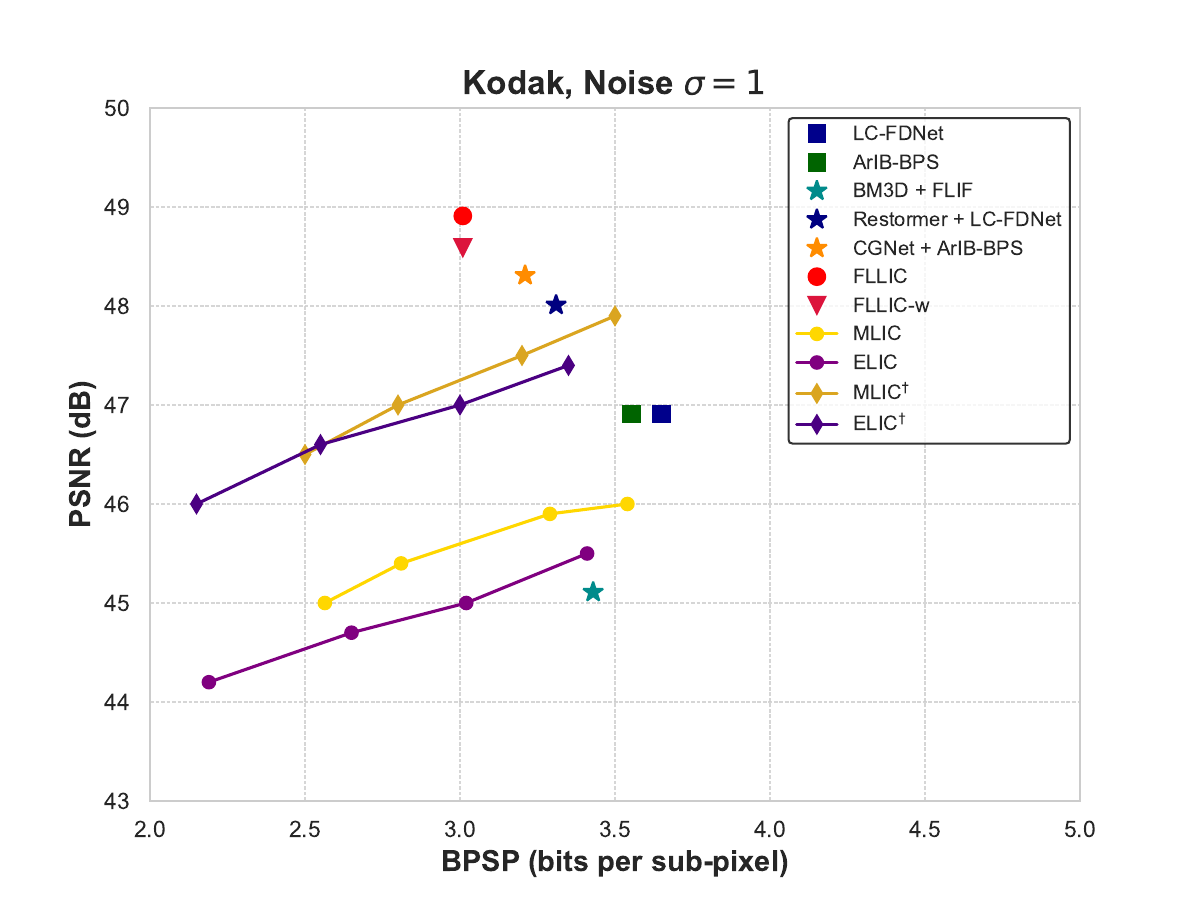}
    \end{minipage}
    \begin{minipage}{0.329\linewidth}
        \includegraphics[width=\linewidth]{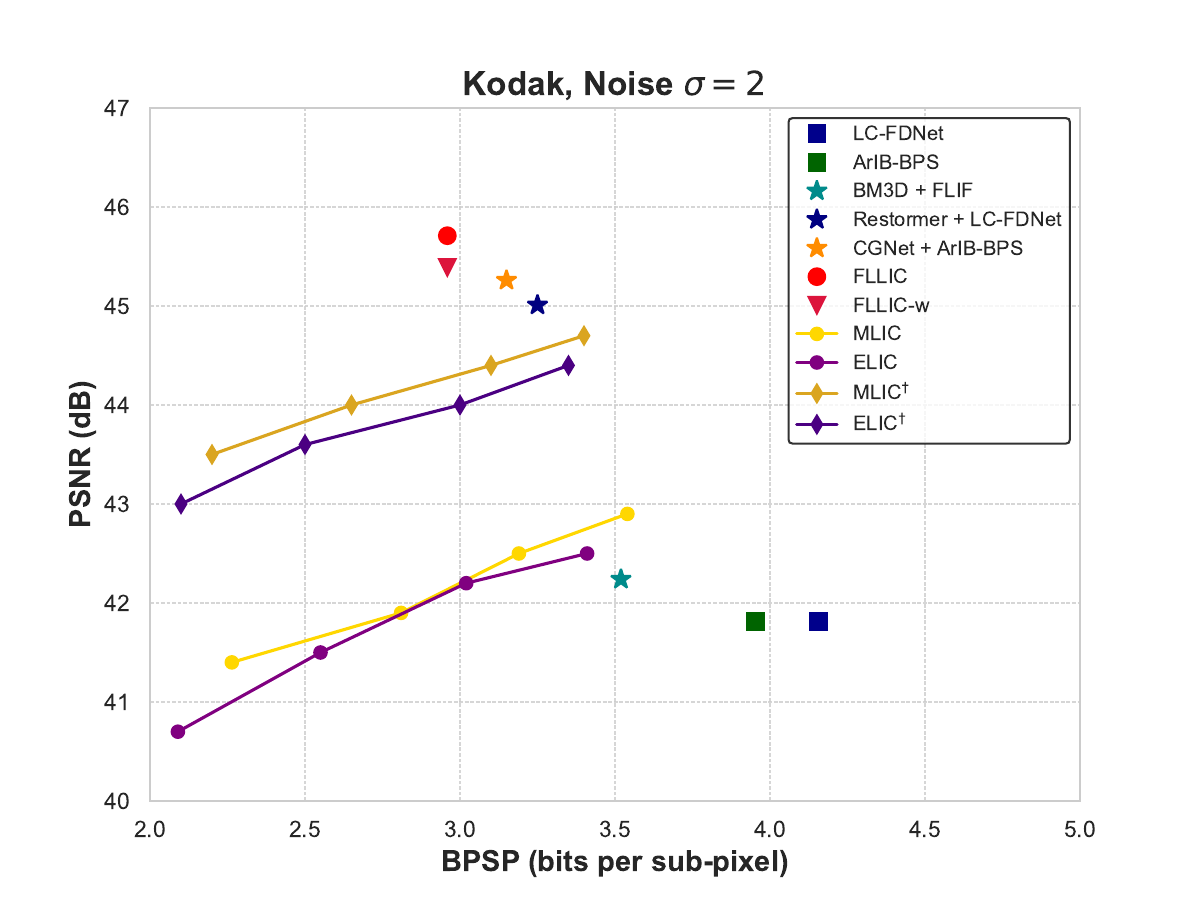}
    \end{minipage}
    \begin{minipage}{0.329\linewidth}
        \includegraphics[width=\linewidth]{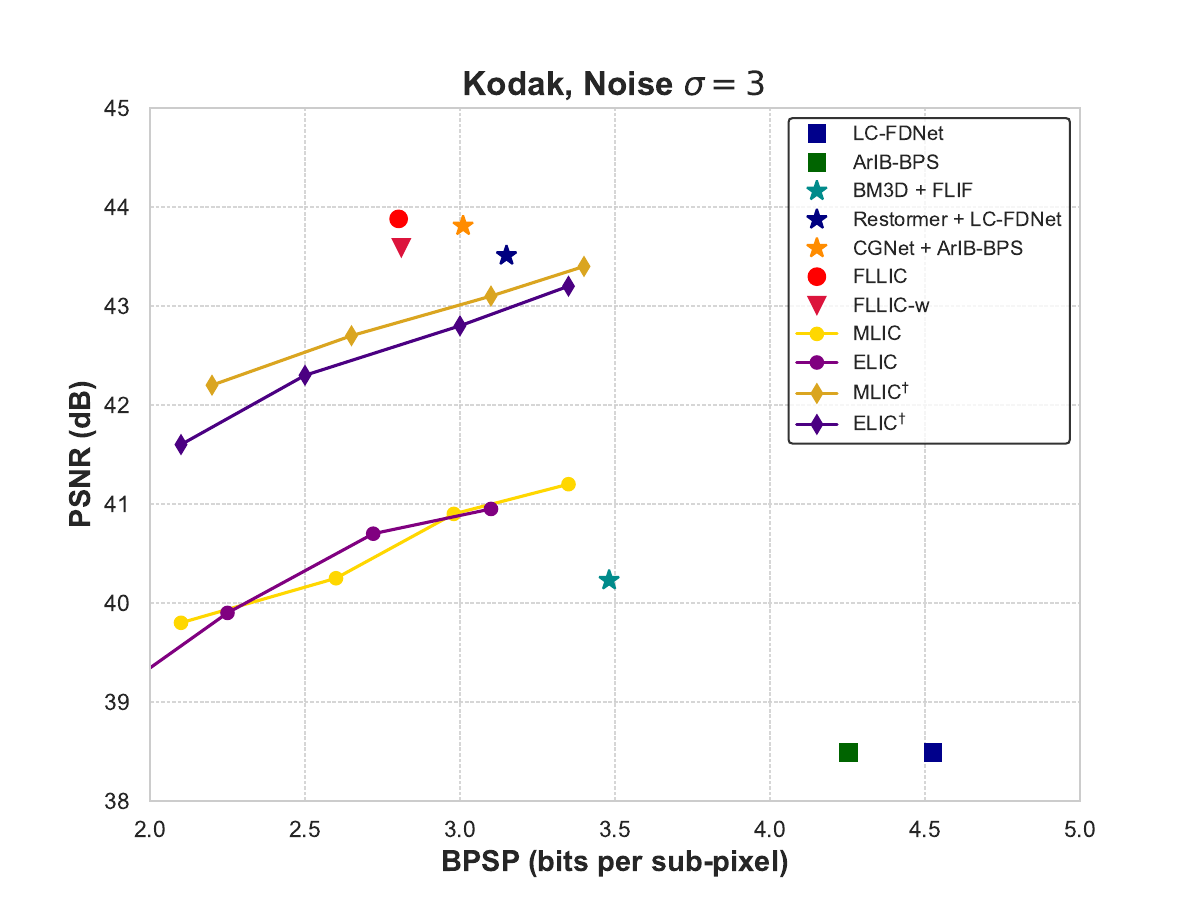}
    \end{minipage} \\
    \begin{minipage}{0.329\linewidth}
        \includegraphics[width=\linewidth]{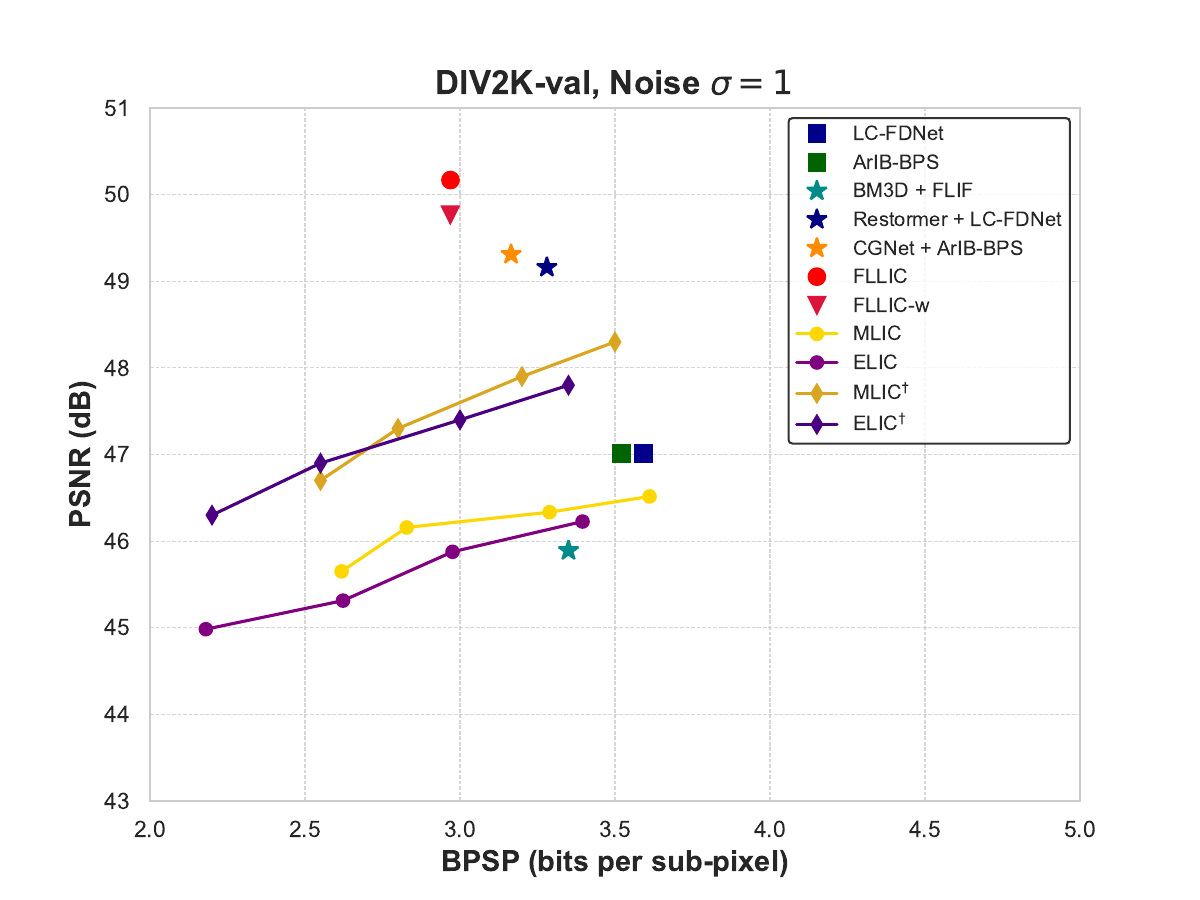}
    \end{minipage}
    \begin{minipage}{0.329\linewidth}
        \includegraphics[width=\linewidth]{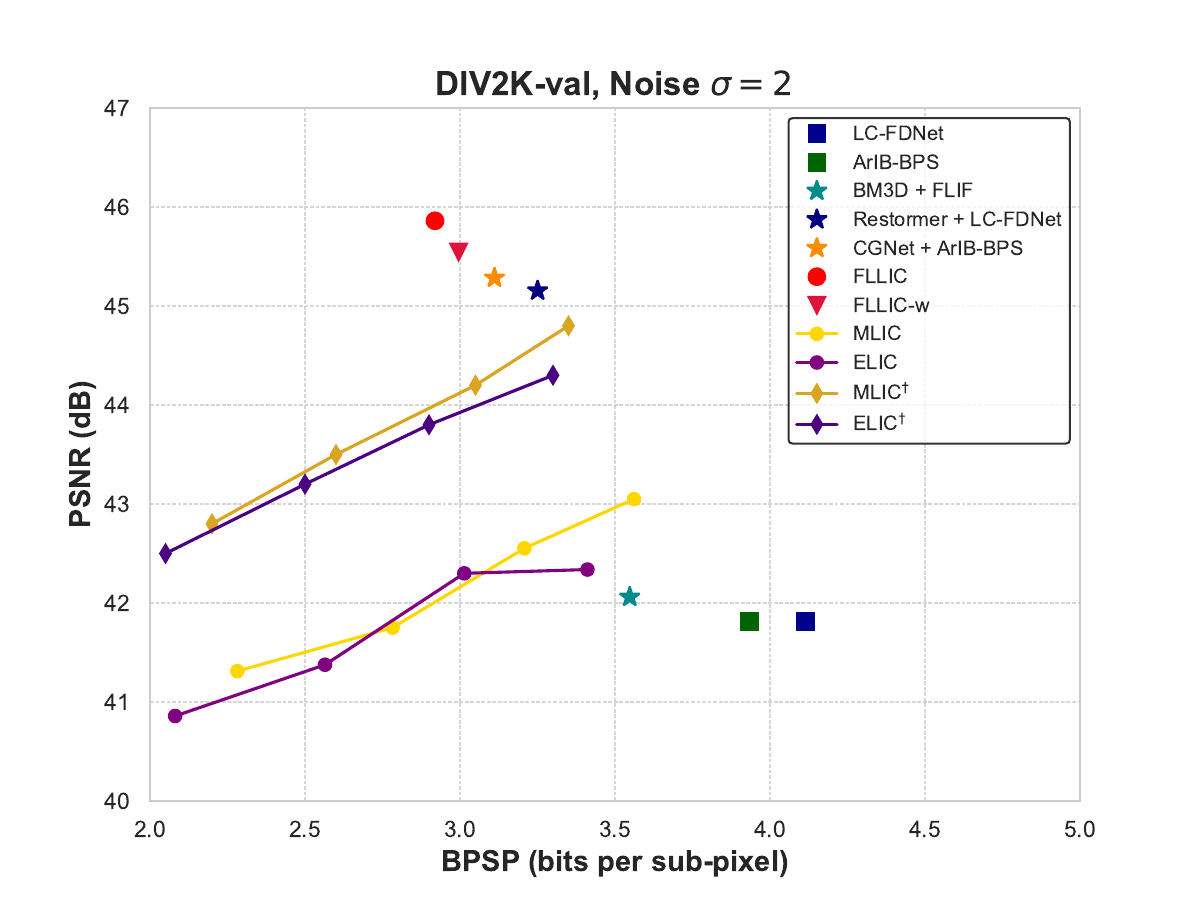}
    \end{minipage}
    \begin{minipage}{0.329\linewidth}
        \includegraphics[width=\linewidth]{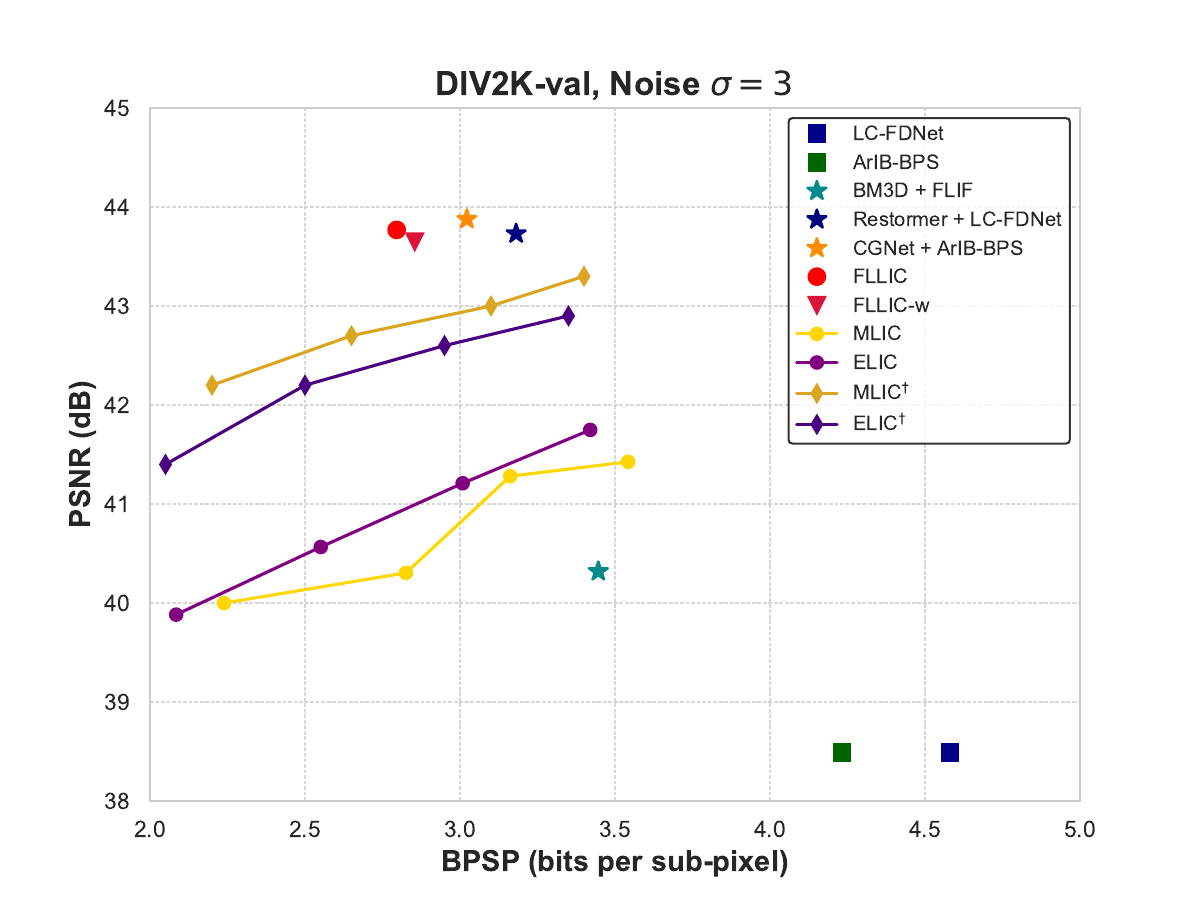}
    \end{minipage} \\
    \caption{
    Rate–Distortion performance comparison of different methods under multiple noise levels ($\sigma = 1, 2, 3$) on the Kodak and DIV2K-val datasets. 
    The x-axis represents bits per sub-pixel (BPSP), and the y-axis denotes peak signal-to-noise ratio (PSNR) in dB. 
    The results highlight the superior joint denoising and compression capability of FLLIC and FLLIC-w, especially under low and moderate noise scenarios.
    In the legend, 
    $\mdblksquare$ denotes lossless compression, 
    $\bigstar$ denotes cascaded denoising + lossless compression, 
    \tikzcircle[black, fill=black]{2.5pt} denotes FLLIC, 
    \textbf{$\blacktriangledown$} denotes FLLIC-w, 
    and $\raisebox{0.4ex}{\rule{0.15cm}{0.4mm}} \tikzcircle[black, fill=black]{2pt} \raisebox{0.4ex}{\rule{0.15cm}{0.4mm}}$ denotes lossy compression.
    MLIC and ELIC correspond to models trained on noisy–noisy image pairs, while the variants MLIC$^{\dagger}$ and ELIC$^{\dagger}$ are trained on noisy–clean pairs to align with FLLIC.
    }
    \label{fig:plot_synthetic}
\end{figure*}
\subsection{Results on Synthetic Datasets}
Based on the results shown in~\autoref{fig:plot_synthetic}, we can draw several key observations regarding the rate-distortion (R-D) performance of various compression methods across different noise levels ($\sigma = 1, 2, 3$) on the Kodak and DIV2K-val datasets.
The proposed FLLIC method consistently achieves the highest PSNR (Peak Signal-to-Noise Ratio) at any given BPSP (bits per sub-pixel) across all noise levels ($\sigma=1,2,3$). It consistently outperforms all other methods in both the Kodak and DIV2K-val datasets, demonstrating superior joint denoising and compression capabilities. FLLIC is especially effective in higher noise scenarios ($\sigma=2$ and $3$), where it maintains better compression efficiency and higher PSNR compared to other methods, reflecting its ability to effectively remove noise while maintaining high fidelity in the reconstruction.

The FLLIC-w method (weakly-supervised version of FLLIC) performs similarly to FLLIC but with slightly lower PSNR values. This difference is more noticeable in low noise scenarios ($\sigma=1$), where FLLIC outperforms FLLIC-w. The performance gap narrows as the noise level increases, indicating that the weak supervision of FLLIC-w is less detrimental in noisier conditions. Despite the slight performance gap, FLLIC-w still offers competitive R-D performance, making it a viable option when fully supervised data is not available.

The Restormer + LC-FDNet approach, which combines the Restormer denoising model with the LC-FDNet lossless compression model, shows strong performance, especially in low noise scenarios ($\sigma = 1$). However, its performance deteriorates with higher noise levels ($\sigma=2$ and $3$), where its PSNR starts to lag behind FLLIC and FLLIC-w. This suggests that while the cascaded denoising and compression strategy can work well in cleaner images, it struggles to match the performance of FLLIC in noisier scenarios due to limitations in noise suppression and coding efficiency. CGNet + ArIB-BPS, which combines CGNet with the ArIB-BPS method, also performs well, but like the Restormer + LC-FDNet, it is more effective in low noise scenarios ($\sigma=1$) and shows a more significant performance drop as noise levels increase. The performance of this approach remains competitive with FLLIC and FLLIC-w in less noisy conditions but fails to match the latter in moderate and high noise scenarios.

The MLIC and ELIC methods, both based on learned lossy compression techniques, perform significantly worse than the lossless and near-lossless methods (such as FLLIC and FLLIC-w) across all noise levels. MLIC achieves higher PSNR than ELIC, but both lag far behind the proposed FLLIC methods. This is expected, as lossy compression methods inherently trade off reconstruction accuracy for compression efficiency, which results in a lower PSNR for the same BPSP.

The MLIC and ELIC methods, both based on learned lossy compression techniques, originally performed significantly worse than the lossless and near-lossless methods (such as FLLIC and FLLIC-w) across all noise levels. This is expected, as standard lossy compression methods inherently prioritize rate–distortion trade-offs with respect to the noisy input, often sacrificing reconstruction fidelity to improve bitrate.
To ensure a fairer comparison under the functionally lossless evaluation protocol, we retrained both MLIC and ELIC using noisy–clean image pairs, aligning their objectives with those of FLLIC. The retrained variants, denoted as MLIC$^\dagger$ and ELIC$^\dagger$, show noticeable improvements in PSNR compared to their original versions. 
However, despite these improvements, both retrained variants still fall far behind FLLIC in terms of rate–distortion performance, requiring substantially higher bitrates yet failing to match FLLIC’s reconstruction quality.

\begin{figure}
    \centering
    \includegraphics[width=0.9\linewidth]{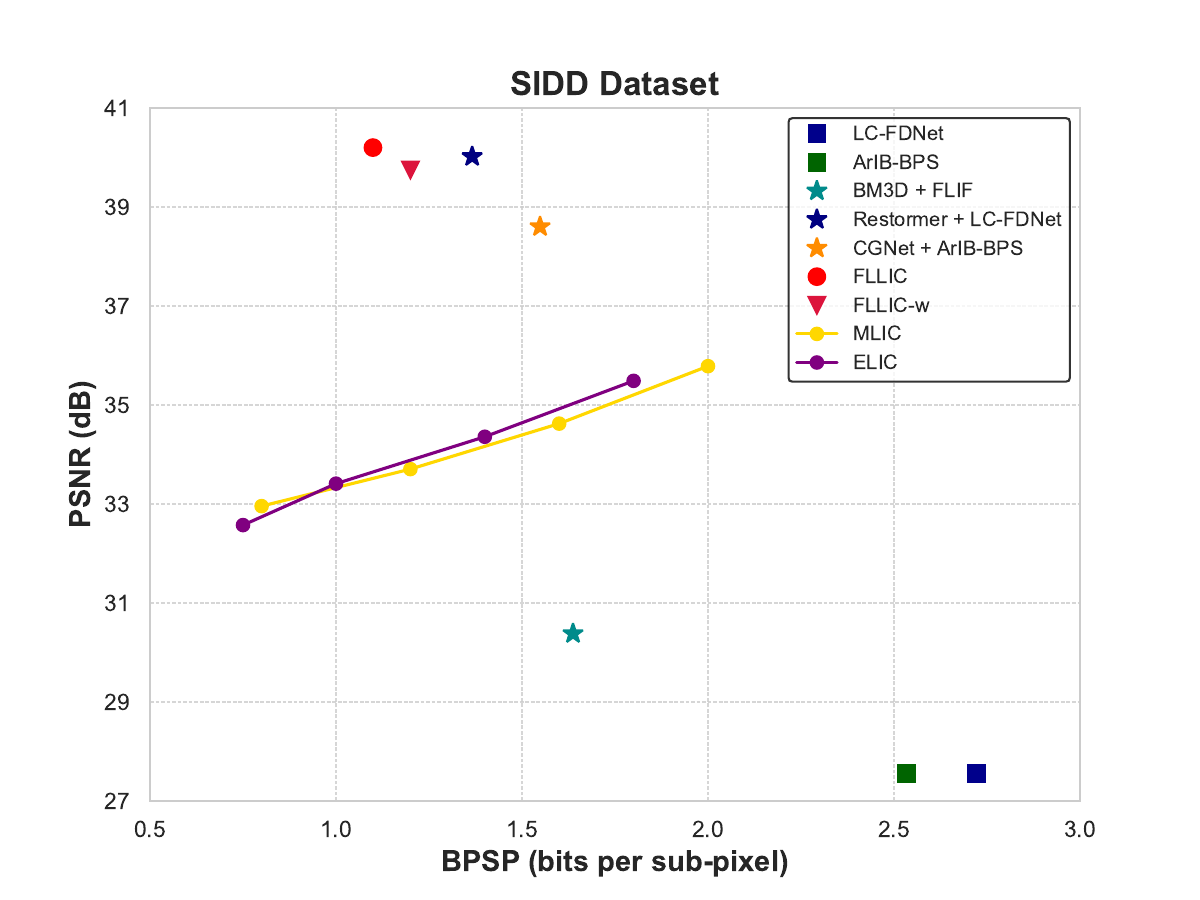}
    \caption{Rate-Distortion performance comparison of different methods on the real-world SIDD dataset. 
    In the legend, $\mdblksquare$ represents lossless compression, $\bigstar$ represents cascaded denoising + lossless compression, 
    \tikzcircle[black, fill=black]{2.5pt} represents FLLIC, \textbf{$\blacktriangledown$} represents FLLIC-w, 
    and $\raisebox{0.4ex}{\rule{0.15cm}{0.4mm}} \tikzcircle[black, fill=black]{2pt} \raisebox{0.4ex}{\rule{0.15cm}{0.4mm}}$ represents lossy compression.}
    \label{fig:plot_sidd}
\end{figure}

\subsection{Results on Real-world Dataset}
\autoref{fig:plot_sidd} presents the Rate-Distortion (R-D) performance comparison of various methods on the real-world SIDD dataset. The results demonstrate the efficacy of the proposed FLLIC method in comparison to a variety of denoising and compression approaches under realistic noise conditions.
The proposed FLLIC method stands out by achieving the highest PSNR values across all BPSP levels. Notably, it outperforms all other methods, including lossy compression methods (ELIC and MLIC), as well as cascaded denoising + lossless compression techniques. The results highlight FLLIC's exceptional ability to simultaneously denoise and compress images, effectively mitigating real-world noise in the process.

FLLIC-w, the weakly-supervised version of FLLIC, follows closely behind but shows a slight decrease in performance compared to the fully-supervised FLLIC. This indicates that the weak supervision does not significantly degrade the performance, especially at higher BPSP values, making FLLIC-w a viable option when fully supervised training data is unavailable.
The Restormer + LC-FDNet approach, which combines transformer-based denoising with lossless compression, demonstrates competitive performance in comparison to FLLIC, particularly at lower BPSP values. However, it still lags behind FLLIC in higher BPSP regions. Similarly, CGNet + ArIB-BPS performs well but struggles to match the performance of FLLIC in this dataset. These findings suggest that while cascaded denoising and compression methods can offer promising results, they are limited by their ability to jointly optimize denoising and compression simultaneously.

In terms of lossy compression, both MLIC and ELIC perform notably worse than FLLIC, especially at higher BPSP values. MLIC and ELIC are optimized for lossy compression, but they do not leverage the denoising aspect that is crucial for real-world noisy images. MLIC and ELIC, while improving PSNR over traditional lossy compression methods, still fall behind FLLIC due to its inability to handle noise as effectively.

\begin{figure*}[!t]
    \centering
    \begin{minipage}{0.329\linewidth}
        \includegraphics[width=\linewidth]{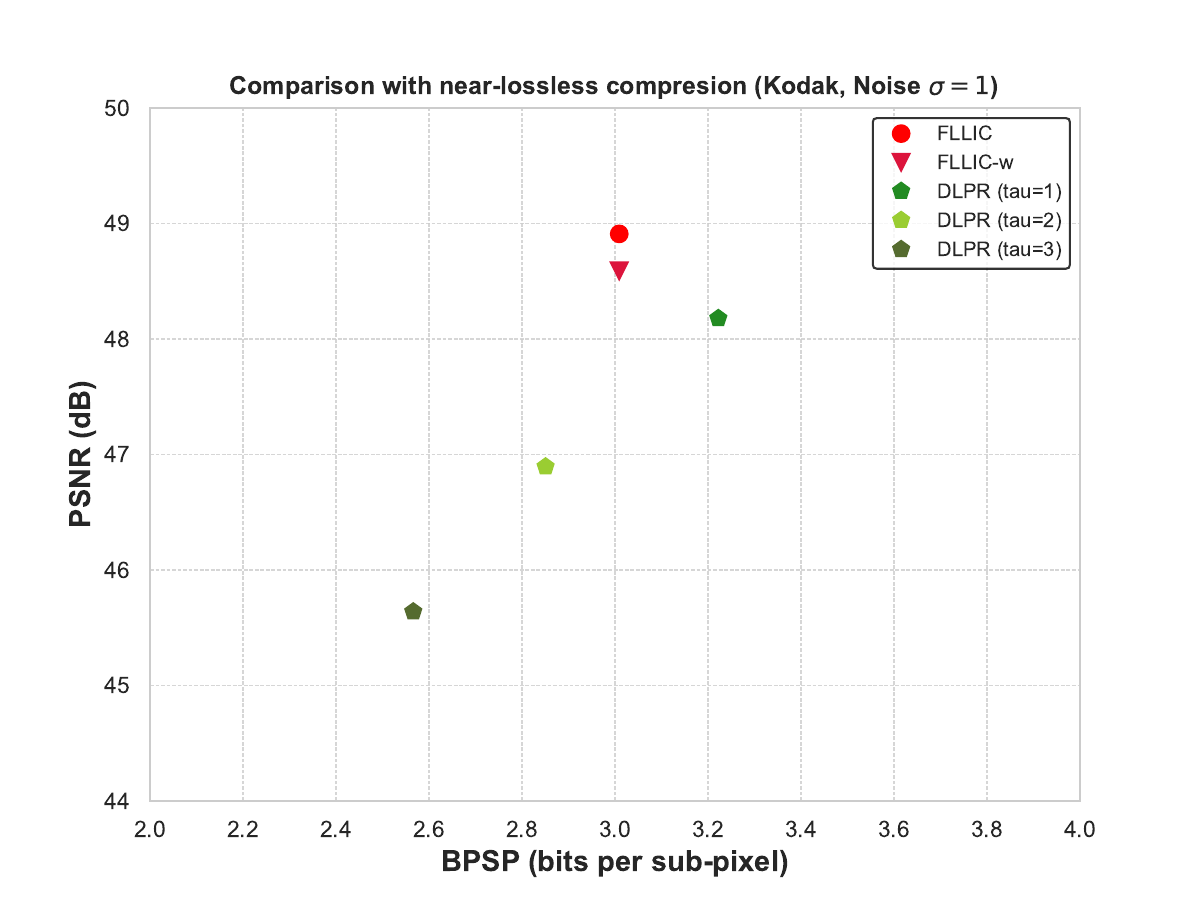}
    \end{minipage}
    \begin{minipage}{0.329\linewidth}
        \includegraphics[width=\linewidth]{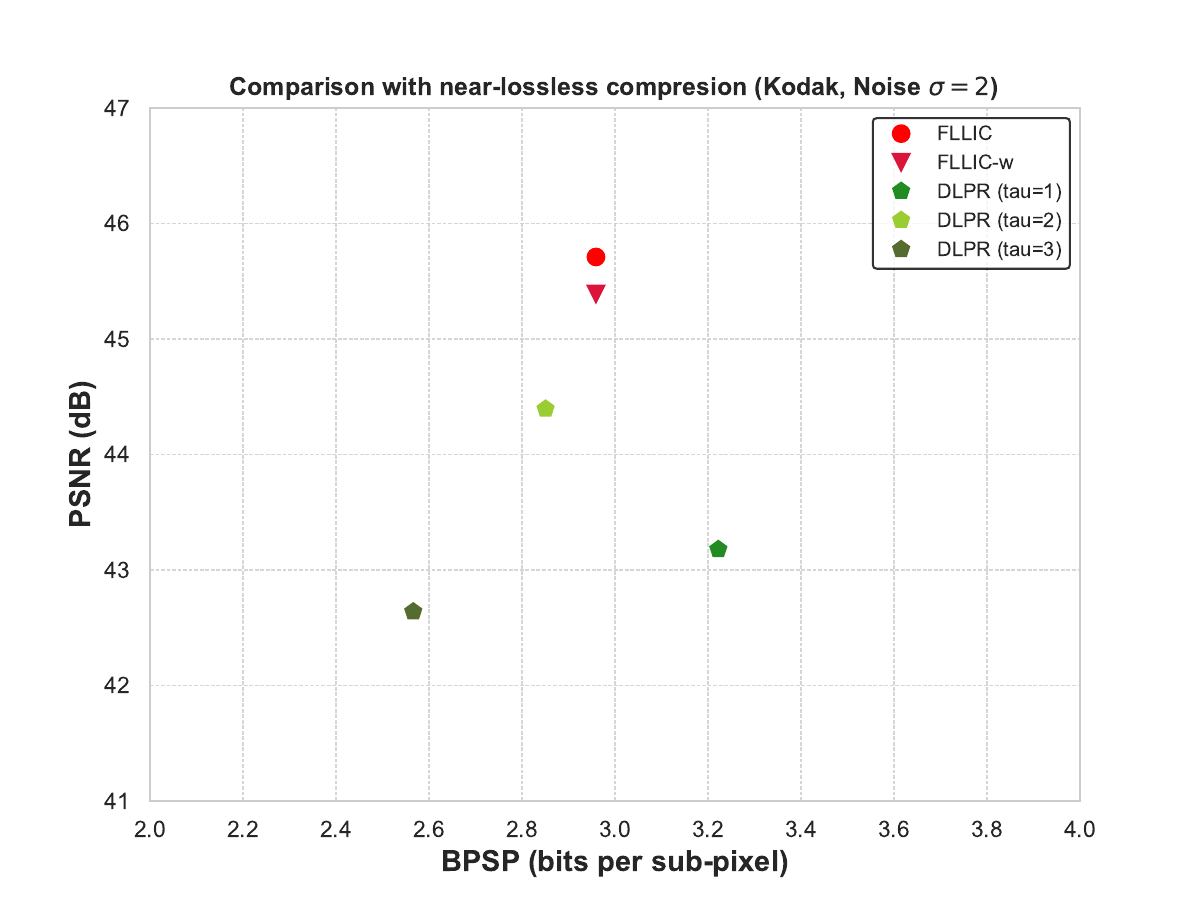}
    \end{minipage}
    \begin{minipage}{0.329\linewidth}
        \includegraphics[width=\linewidth]{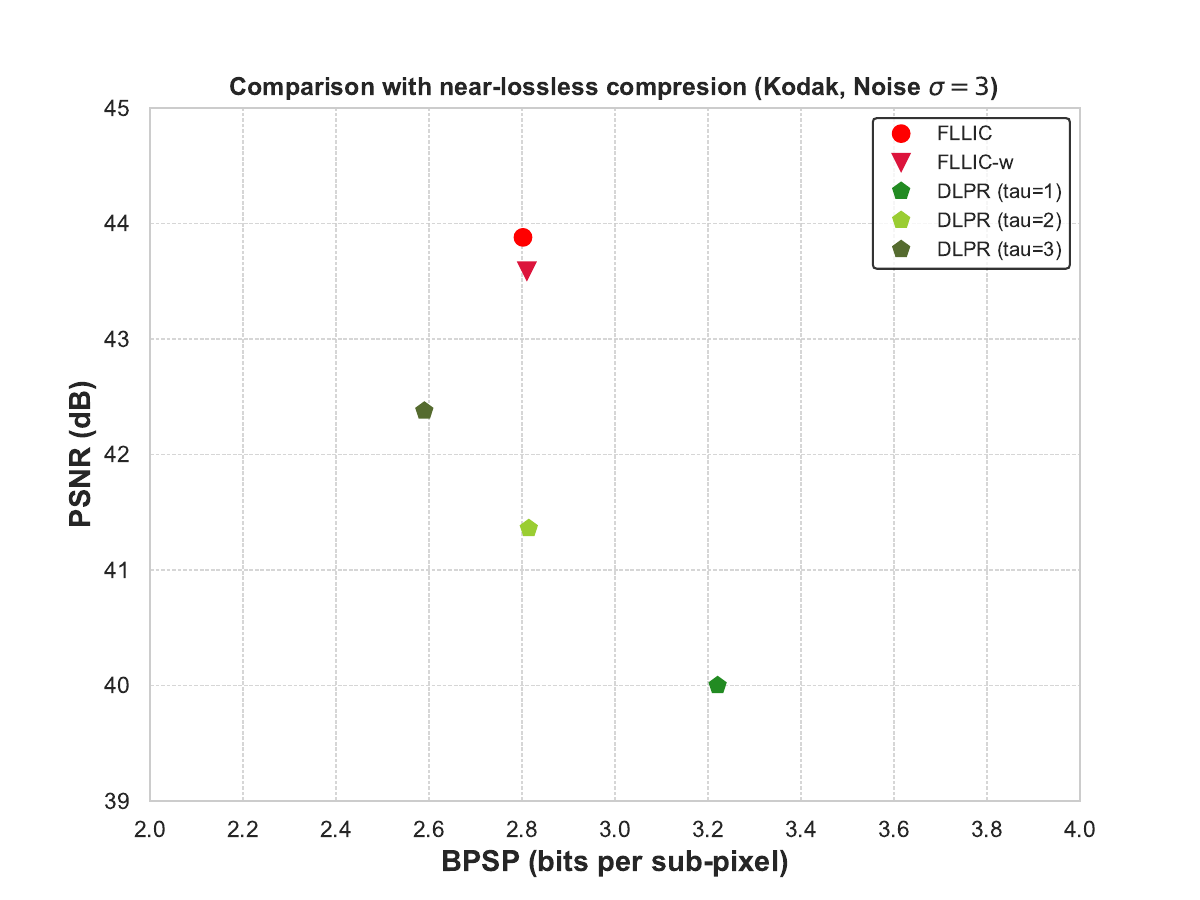}
    \end{minipage} \\
    \caption{Rate-Distortion performance comparison of FLLIC, FLLIC-w, and DLPR (near-lossless) methods on Kodak dataset with varying noise levels ($\sigma = 1,2,3$). The x-axis represents the bits per sub-pixel (BPSP), and the y-axis denotes the peak signal-to-noise ratio (PSNR) in dB. The plots show the performance of FLLIC and its weakly-supervised version (FLLIC-w) alongside the near-lossless compression method DLPR for three different values of $\tau$.
    }
    \label{fig:plot_nl}
\end{figure*}
\subsection{Comparison with Near-lossless Compression}
\autoref{fig:plot_nl} shows the Rate-Distortion (R-D) performance comparison of FLLIC, FLLIC-w, and DLPR (near-lossless compression) methods on the Kodak dataset, with varying noise levels ($\sigma = 1, 2, 3$). The plots compare the performance of FLLIC and its weakly-supervised version (FLLIC-w) with the near-lossless compression method DLPR for three different values of noise level $\tau$.

In all noise scenarios, FLLIC provides the best performance in terms of PSNR, followed by FLLIC-w. Specifically,
at noise level $\sigma = 1$ (first plot), FLLIC achieves the highest PSNR (~49 dB) at a BPSP of around 3.0, with FLLIC-w performing slightly lower (~48.5 dB). The near-lossless compression method DLPR at $\tau = 1$ shows a performance drop (~47.8 dB), and as $\tau$ increases, its performance further declines, with $\tau = 3$ reaching the lowest PSNR (~45 dB). At $\sigma = 2$ (second plot), the performance of all methods decreases as noise increases, but FLLIC still outperforms the others, achieving PSNR of around 46.5 dB at a BPSP of 2.8. DLPR’s performance at $\tau = 1$ continues to drop (~44 dB), and the gap between FLLIC and DLPR widens as $\tau$ increases. At $\sigma = 3$ (third plot), FLLIC remains the best-performing method (~45 dB), while DLPR’s PSNR continues to degrade, reaching as low as 40 dB at $\tau = 3$. These results demonstrate FLLIC’s superior performance and robustness, especially in low-noise scenarios and at higher noise levels, compared to DLPR.

\begin{table*}[t]
  \centering
  \caption{Inference time (ms) of various methods for encoding and decoding a $512\times512$ image on a Nvidia RTX 4090 GPU. The methods include cascaded denoising + lossless compression, lossy compression, near-lossless compression and the proposed FLLIC method.}
  \label{tab:time}
  \renewcommand\arraystretch{1.0}
  \begin{tabular}{c|cc|cc|c|c|c|c|c|c}
  \toprule
   & \multicolumn{4}{c|}{Cascaded} & \multicolumn{2}{c|}{Lossy} & \multicolumn{2}{c|}{Near-lossless} & \multicolumn{2}{c}{Proposed FLLIC} \\
  \midrule
   & Restormer & LC-FDNet & CGNet & ArIB-BPS & ELIC & MLIC & \multicolumn{2}{c|}{DLPR} & FLLIC & FLLIC-w \\
  \midrule
  Encoding & 465 ms & 428 ms & 510 ms & 520 ms & 380 ms & 497 ms & \multicolumn{2}{c|}{1227ms} & 124 ms & 181 ms \\
  \midrule
  Decoding & - & 462 ms & - & 495 ms & 235 ms & 312 ms & \multicolumn{2}{c|}{2518ms} & 119 ms & 165 ms \\
  \bottomrule
  \end{tabular}
\end{table*}

\subsection{Inference Time}
We measure the inference time required for encoding and decoding a $512\times512$ image on a Nvidia RTX 4090 GPU. The detailed inference time of competing methods is listed in~\autoref{tab:time}. For the cascaded method, which involves denoising followed by lossless compression, the encoding step takes about 428 ms for Restormer + LC-FDNet, and about 510 ms for CGNet, with decoding times of 462 ms and 495 ms, respectively. In comparison, the proposed FLLIC method requires only 124 ms for encoding and 119 ms for decoding, which is significantly faster than the cascaded methods by an order of magnitude. Additionally, the weakly-supervised version, FLLIC-w, performs slightly slower, with encoding taking 181 ms and decoding taking 165 ms. 
Despite its speed advantage, FLLIC still achieves superior rate-distortion performance, highlighting the efficiency and effectiveness of the proposed approach.

\subsection{Visualization of Noise-Adaptive Quantization Maps}
\label{subsec:quant-map-vis}

Understanding how noise statistics influence the learned content-adaptive quantization maps is important for interpreting the behavior of our proposed FLLIC framework. In this subsection, we provide both a rationale for the design and empirical visualizations to clarify this mechanism.

\paragraph{Design rationale.}
It is intractable to derive an explicit, closed-form mapping from local noise characteristics to optimal quantization step sizes. First, the relevant statistics are determined jointly by the underlying clean signal and the spatially varying noise power. Second, the distributions of latent and hyper-latent representations after nonlinear analysis transforms are highly signal-dependent. 
To overcome these challenges, FLLIC leverages the estimated clean entropy $H(I)$ as an informative proxy that implicitly captures the effect of noise on local compressibility. 
% Specifically, higher noise levels tend to increase the entropy gap between the noisy input and the estimated clean image, leading to smaller predicted $H(I)$ values in noise-dominated regions. 
By conditioning the quantization step prediction networks on $H(I)$, the model learns to allocate larger quantization steps in regions with higher noise and finer steps in regions rich in structural content. This adaptive behavior emerges automatically through end-to-end training and does not require handcrafted analytical formulas.

\begin{figure}[t]
  \centering
  \includegraphics[width=\linewidth]{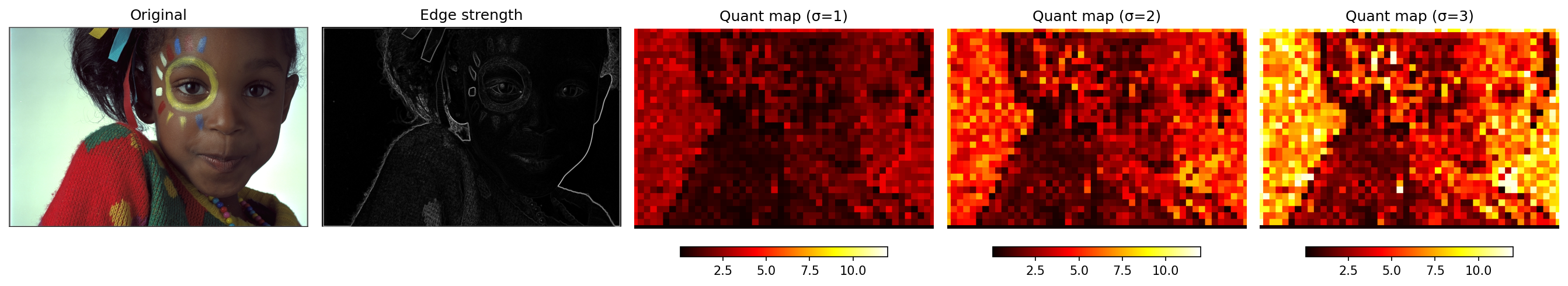} \\
  \includegraphics[width=\linewidth]{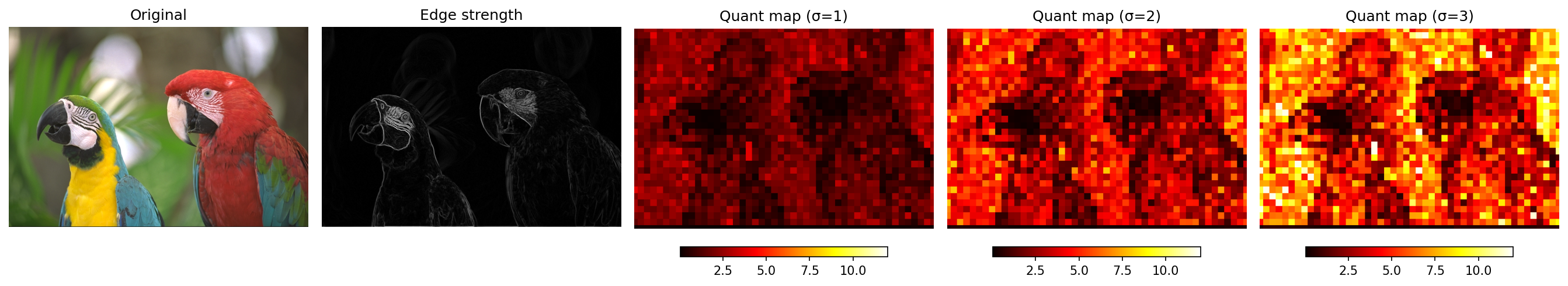}
  \caption{Visualization of learned quantization maps ($Q_y$ and $Q_z$) for four example images under varying noise levels. Each row corresponds to a different input image, and within each, quantization step sizes increase with global noise level while retaining finer steps near structural edges.}
  \label{fig:quant-maps}
\end{figure}

\paragraph{Visualization results.}
To provide empirical evidence of this learned behavior, we visualize the predicted $Q_y$ (for main latents) and $Q_z$ (for hyper-latents) maps under different global noise levels. Specifically, we use the same test image corrupted with uniform Gaussian noise at three standard deviations $\sigma = 1, 2, 3$, and plot the corresponding heatmaps of the predicted quantization steps.
\autoref{fig:quant-maps} shows two key trends. First, as the \emph{global} noise level increases, the quantization steps overall become larger, reflecting the reduced need to preserve fine details at lower signal-to-noise ratios. Second, within each noise level, the maps consistently assign finer steps along structural edges and coarser steps in smoother areas, indicating that the model also adapts to local image content. These observations confirm that our learned quantization maps respond simultaneously to global noise statistics and local structural importance.

\subsection{Factorized vs. Auto-Regressive Entropy Models}
\label{subsec:factor-vs-AR}

In our framework, the entropy model for the hyperlatents $z$ adopts a fully factorized form. This design choice is widely adopted in learned image compression due to the typically minor contribution of hyperlatents to the overall bitrate and the desire to maintain fast, parallel decoding. In contrast, complex context models such as auto-regressive masked convolutions can offer marginal compression gains at the cost of significantly increased computational overhead.

To empirically validate this design choice, we additionally implemented an auto-regressive variant of FLLIC (denoted as FLLIC-A) that employs masked convolutions to capture spatial dependencies within $z$. Both versions were evaluated on the Kodak dataset under AWGN noise levels $\sigma = 1, 2, 3$. The results are summarized in \autoref{tab:entropy_model}.

\begin{table}[t]
\centering
\caption{Comparison between factorized (FLLIC) and auto-regressive (FLLIC-A) entropy models on Kodak. Reported: BPP / PSNR (dB) / Inference time change.}
\renewcommand\arraystretch{1.0}
\begin{tabular}{c|cc|c|c|c}
\toprule
$\sigma$ & FLLIC & FLLIC-A & $\Delta$ BPP & $\Delta$ PSNR & $\Delta$ Time \\
\midrule
1 & 3.009 / 48.91 & 2.960 / 49.03 & $-1.6\%$ & $+0.12$ & $+105\%$ \\
2 & 2.959 / 45.71 & 2.915 / 45.82 & $-1.5\%$ & $+0.11$ & $+113\%$ \\
3 & 2.802 / 43.88 & 2.760 / 44.04 & $-1.5\%$ & $+0.16$ & $+120\%$ \\
\bottomrule
\end{tabular}
\label{tab:entropy_model}
\end{table}

As shown, the auto-regressive model provides only marginal bitrate reduction ($\sim\!1.5\%$) and minor PSNR gains ($<0.2$ dB), while incurring over $100\%$ increase in inference latency. Considering the negligible compression improvement but substantial runtime penalty, we retain the factorized entropy model as the default for its superior balance between efficiency and performance in practical deployment.

\subsection{Ablation on Entropy Estimation and Adaptive Quantization}
\label{subsec:ablation_entropy_quant}

To evaluate the contribution of key modules in FLLIC, we conduct an ablation study on the Kodak dataset with synthetic AWGN noise ($\sigma = 1, 2, 3$). Specifically, we progressively disable the entropy estimation branch $H(I)$ and the content-adaptive quantization modules $Q_y, Q_z$. The results are tabulated in \autoref{tab:ablation_study} and discussed below:
(i) Removing $H(I)$ and replacing it with a global scalar increases the bitrate by $6.8\%$ on average and causes a slight but consistent PSNR drop ($\leq 0.1$\,dB). This confirms that entropy estimation provides a strong global prior to reduce redundancy.
(ii) Disabling spatially adaptive $Q_y,Q_z$ further increases bitrate by $9.2\%$ and leads to slightly larger PSNR degradation ($\sim\!0.1$\,dB), indicating that local adaptation is essential for handling heterogeneous content.
(iii) Removing both components causes over $15\%$ bitrate deterioration and up to $0.18$\,dB loss in PSNR, highlighting their complementary roles.

\begin{table}[t]
\centering
\caption{Ablation study on entropy estimation and adaptive quantization. Reported: BPP / PSNR (dB).}
\begin{tabular}{c|c|c|c|c}
\toprule
$\sigma$ & FLLIC (full) & w/o $H(I)$ & w/o $Q_y,Q_z$ & w/o Both \\
\midrule
1 & 3.009 / 48.91 & 3.214 / 48.85 & 3.285 / 48.82 & 3.476 / 48.77 \\
2 & 2.959 / 45.71 & 3.160 / 45.66 & 3.230 / 45.62 & 3.417 / 45.55 \\
3 & 2.802 / 43.88 & 2.993 / 43.82 & 3.061 / 43.78 & 3.236 / 43.70 \\
\bottomrule
\end{tabular}
\label{tab:ablation_study}
\end{table}

\subsection{Impact of Noise Variance Estimation in Entropy Modeling}
\label{subsec:entropy_sigma_ablation}

In our main FLLIC design, the clean-entropy estimation branch takes the noise standard deviation $\sigma_N$ as input to inform its prediction of $H(I)$. While this was initially studied under oracle conditions with ground-truth $\sigma_N$ provided, we further assess the impact of replacing this oracle information with either (i) the estimated $\widehat{\sigma}_N$ from the NLE or (ii) removing $\sigma_N$ altogether.
\autoref{tab:nle_ablation} reports the comparative results. Using the estimated $\widehat{\sigma}_N$ leads to only marginal bitrate overhead ($\leq 1.36\%$) compared to the ground-truth, confirming the effectiveness of the NLE. Omitting $\sigma_N$ entirely causes a larger bitrate increase (up to $7.18\%$), demonstrating that providing even an estimated noise level is beneficial. Nonetheless, FLLIC without $\sigma_N$ input still outperforms cascaded baselines, verifying its robustness.

\begin{table}[t]
\centering
\caption{Impact of ground-truth vs.\ estimated vs.\ absent noise variance on rate--distortion performance. BPP / PSNR (dB).}
\label{tab:nle_ablation}
\renewcommand{\arraystretch}{1.0}
\begin{tabular}{c|c|c|c}
\toprule
$\sigma$ & GT $\sigma_N$ & Est.\ $\widehat{\sigma}_N$ & No $\sigma_N$ \\
\midrule
1 & 3.009 / 48.91 & 3.050 / 48.90 ($+1.35\%$) & 3.225 / 48.78 ($+7.18\%$) \\
2 & 2.959 / 45.71 & 2.990 / 45.70 ($+1.05\%$) & 3.150 / 45.60 ($+6.45\%$) \\
3 & 2.802 / 43.88 & 2.840 / 43.85 ($+1.36\%$) & 2.960 / 43.72 ($+5.64\%$) \\
\bottomrule
\end{tabular}
\end{table}

\subsection{Noise-Adaptive Extension of FLLIC}
\label{subsec:noise_adaptive}
In the main experiments, FLLIC was trained with a separate model for each known noise level ($\sigma$) to assess its performance upper bound under controlled conditions. While this design facilitates fair benchmarking, it is not realistic for real-world scenarios where noise characteristics are often unknown or time-varying. To address this limitation, we propose a noise-adaptive variant of FLLIC that generalizes to unseen noise levels without requiring retraining.

The noise-adaptive FLLIC augments the original model with two lightweight components:
(i) \textit{Noise-Level Estimator (NLE).} A four-layer depthwise–separable CNN ($<\!0.15$M parameters) that predicts the global noise standard deviation $\widehat{\sigma}_N$ from the noisy input. The NLE is trained jointly with the codec, requiring no additional supervision; (ii) \textit{Conditional Modulation.} The predicted $\widehat{\sigma}_N$ is injected into the entropy-estimation and quantization branches ($H(I)$, $Q_y$, $Q_z$) via FiLM-style affine modulation, enabling dynamic adjustment of coding strategies according to the estimated noise.

We train a single adaptive FLLIC model on mixed AWGN data ($\sigma=1\!-\!5$) and evaluate it on unseen noise levels ($\sigma=1.5, 3.5, 4$). The results in \autoref{tab:adaptive_fllic} show:
(i) Compared to expert per-noise models, the adaptive model incurs only a $1.93\%$ average bitrate overhead under oracle $\sigma$; (ii) When relying on the NLE prediction, the additional penalty remains below $0.8\%$, verifying the estimator's accuracy.
Besides, the NLE adds only 3.6\,ms latency per $512\!\times\!512$ image on RTX 4090, with no extra memory at decoding.

\begin{table}[t]
\centering
\caption{Evaluation of the noise-adaptive FLLIC variant. Reported: BPP / PSNR (dB) / Overhead.}
\resizebox{\linewidth}{!}{
\begin{tabular}{c|c|c|c}
\toprule
$\sigma$  & Expert FLLIC & Adaptive (Oracle) & Adaptive (Estimator) \\
\midrule
$\sigma=1.5$ & 2.980 / 47.20 & 3.035 / 47.22 ($+1.85\%$) & 3.055 / 47.23 ($+2.52\%$) \\
$\sigma=3.5$ & 2.810 / 43.10 & 2.865 / 43.12 ($+1.96\%$) & 2.880 / 43.14 ($+2.49\%$) \\
$\sigma=4$   & 2.780 / 42.30 & 2.835 / 42.33 ($+1.98\%$) & 2.850 / 42.35 ($+2.52\%$) \\
\midrule
\multicolumn{2}{l|}{\textbf{Bitrate overhead}} & $+1.93\%$ & $<0.8\%$ vs Oracle \\
\bottomrule
\end{tabular}
}
\label{tab:adaptive_fllic}
\end{table}

\section{Conclusion}
We introduce a new paradigm called functionally lossless image compression (FLLIC), which integrates the two tasks of denoising and compression. FLLIC engages in lossless/near-lossless compression of optimally denoised images, with optimality tailored to specific tasks. While not strictly adhering to the literal meaning of losslessness concerning the noisy input, FLLIC aspires to achieve the optimal reconstruction of the latent noise-free original image.
Extensive empirical investigations underscore the state-of-the-art performance of FLLIC in the realm of joint denoising and compression for noisy images, concurrently exhibiting advantages in terms of computational efficiency and cost-effectiveness.

%%%%%%%%% REFERENCES

\bibliographystyle{IEEEtran}
\bibliography{fllic}

% Generated by IEEEtran.bst, version: 1.14 (2015/08/26)
\begin{thebibliography}{10}
\providecommand{\url}[1]{#1}
\csname url@samestyle\endcsname
\providecommand{\newblock}{\relax}
\providecommand{\bibinfo}[2]{#2}
\providecommand{\BIBentrySTDinterwordspacing}{\spaceskip=0pt\relax}
\providecommand{\BIBentryALTinterwordstretchfactor}{4}
\providecommand{\BIBentryALTinterwordspacing}{\spaceskip=\fontdimen2\font plus
\BIBentryALTinterwordstretchfactor\fontdimen3\font minus \fontdimen4\font\relax}
\providecommand{\BIBforeignlanguage}[2]{{%
\expandafter\ifx\csname l@#1\endcsname\relax
\typeout{** WARNING: IEEEtran.bst: No hyphenation pattern has been}%
\typeout{** loaded for the language `#1'. Using the pattern for}%
\typeout{** the default language instead.}%
\else
\language=\csname l@#1\endcsname
\fi
#2}}
\providecommand{\BIBdecl}{\relax}
\BIBdecl

\bibitem{Balle2017_End}
J.~Ball{\'{e}}, V.~Laparra, and E.~P. Simoncelli, ``End-to-end optimized image compression,'' in \emph{5th International Conference on Learning Representations, {ICLR}}, 2017.

\bibitem{Theis2017_Lossy}
L.~Theis, W.~Shi, A.~Cunningham, and F.~Husz{\'{a}}r, ``Lossy image compression with compressive autoencoders,'' in \emph{5th International Conference on Learning Representations, {ICLR}}, 2017.

\bibitem{Agustsson2017_Soft}
E.~Agustsson, F.~Mentzer, M.~Tschannen, L.~Cavigelli, R.~Timofte, L.~Benini, and L.~V. Gool, ``Soft-to-hard vector quantization for end-to-end learning compressible representations,'' in \emph{Advances in Neural Information Processing Systems 30}, 2017, pp. 1141--1151.

\bibitem{Balle2018_Vari}
J.~Ball{\'{e}}, D.~Minnen, S.~Singh, S.~J. Hwang, and N.~Johnston, ``Variational image compression with a scale hyperprior,'' in \emph{6th International Conference on Learning Representations, {ICLR}}.\hskip 1em plus 0.5em minus 0.4em\relax OpenReview.net, 2018.

\bibitem{Minnen2018_Joint}
D.~Minnen, J.~Ball{\'{e}}, and G.~Toderici, ``Joint autoregressive and hierarchical priors for learned image compression,'' in \emph{Advances in Neural Information Processing Systems 31}, 2018, pp. 10\,794--10\,803.

\bibitem{Mentzer2018_Cond}
F.~Mentzer, E.~Agustsson, M.~Tschannen, R.~Timofte, and L.~V. Gool, ``Conditional probability models for deep image compression,'' in \emph{2018 {IEEE} Conference on Computer Vision and Pattern Recognition, {CVPR}}, 2018, pp. 4394--4402.

\bibitem{Lee2019_Context}
J.~Lee, S.~Cho, and S.~Beack, ``Context-adaptive entropy model for end-to-end optimized image compression,'' in \emph{7th International Conference on Learning Representations, {ICLR}}, 2019.

\bibitem{Chen2020_Learned}
Z.~Cheng, H.~Sun, M.~Takeuchi, and J.~Katto, ``Learned image compression with discretized gaussian mixture likelihoods and attention modules,'' in \emph{2020 {IEEE/CVF} Conference on Computer Vision and Pattern Recognition, {CVPR}}, 2020, pp. 7936--7945.

\bibitem{hific}
F.~Mentzer, G.~D. Toderici, M.~Tschannen, and E.~Agustsson, ``High-fidelity generative image compression,'' \emph{Advances in Neural Information Processing Systems}, vol.~33, pp. 11\,913--11\,924, 2020.

\bibitem{agdl}
X.~Zhang and X.~Wu, ``Attention-guided image compression by deep reconstruction of compressive sensed saliency skeleton,'' in \emph{Proceedings of the IEEE/CVF Conference on Computer Vision and Pattern Recognition}, 2021, pp. 13\,354--13\,364.

\bibitem{he2021checkerboard}
D.~He, Y.~Zheng, B.~Sun, Y.~Wang, and H.~Qin, ``Checkerboard context model for efficient learned image compression,'' in \emph{Proceedings of the IEEE/CVF Conference on Computer Vision and Pattern Recognition}, 2021, pp. 14\,771--14\,780.

\bibitem{yang2021slimmable}
F.~Yang, L.~Herranz, Y.~Cheng, and M.~G. Mozerov, ``Slimmable compressive autoencoders for practical neural image compression,'' in \emph{Proceedings of the IEEE/CVF Conference on Computer Vision and Pattern Recognition}, 2021, pp. 4998--5007.

\bibitem{kim2022joint}
J.-H. Kim, B.~Heo, and J.-S. Lee, ``Joint global and local hierarchical priors for learned image compression,'' in \emph{Proceedings of the IEEE/CVF Conference on Computer Vision and Pattern Recognition}, 2022, pp. 5992--6001.

\bibitem{he2022elic}
D.~He, Z.~Yang, W.~Peng, R.~Ma, H.~Qin, and Y.~Wang, ``Elic: Efficient learned image compression with unevenly grouped space-channel contextual adaptive coding,'' in \emph{Proceedings of the IEEE/CVF Conference on Computer Vision and Pattern Recognition}, 2022, pp. 5718--5727.

\bibitem{lee2022selective}
J.~Lee, S.~Jeong, and M.~Kim, ``Selective compression learning of latent representations for variable-rate image compression,'' \emph{arXiv preprint arXiv:2211.04104}, 2022.

\bibitem{addl}
X.~Zhang and X.~Wu, ``Dual-layer image compression via adaptive downsampling and spatially varying upconversion,'' \emph{arXiv preprint arXiv:2302.06096}, 2023.

\bibitem{zou2022devil}
R.~Zou, C.~Song, and Z.~Zhang, ``The devil is in the details: Window-based attention for image compression,'' in \emph{Proceedings of the IEEE/CVF conference on computer vision and pattern recognition}, 2022, pp. 17\,492--17\,501.

\bibitem{zhang2023_lvqac}
X.~Zhang and X.~Wu, ``Lvqac: Lattice vector quantization coupled with spatially adaptive companding for efficient learned image compression,'' in \emph{Proceedings of the IEEE/CVF Conference on Computer Vision and Pattern Recognition}, 2023, pp. 10\,239--10\,248.

\bibitem{mentzer2019practical}
F.~Mentzer, E.~Agustsson, M.~Tschannen, R.~Timofte, and L.~V. Gool, ``Practical full resolution learned lossless image compression,'' in \emph{Proceedings of the IEEE/CVF conference on computer vision and pattern recognition}, 2019, pp. 10\,629--10\,638.

\bibitem{mentzer2020learning}
F.~Mentzer, L.~V. Gool, and M.~Tschannen, ``Learning better lossless compression using lossy compression,'' in \emph{Proceedings of the IEEE/CVF Conference on Computer Vision and Pattern Recognition}, 2020, pp. 6638--6647.

\bibitem{kingma2019bit}
F.~Kingma, P.~Abbeel, and J.~Ho, ``Bit-swap: Recursive bits-back coding for lossless compression with hierarchical latent variables,'' in \emph{International Conference on Machine Learning}.\hskip 1em plus 0.5em minus 0.4em\relax PMLR, 2019, pp. 3408--3417.

\bibitem{townsend2019hilloc}
J.~Townsend, T.~Bird, J.~Kunze, and D.~Barber, ``Hilloc: Lossless image compression with hierarchical latent variable models,'' \emph{arXiv preprint arXiv:1912.09953}, 2019.

\bibitem{hoogeboom2019integer}
E.~Hoogeboom, J.~Peters, R.~Van Den~Berg, and M.~Welling, ``Integer discrete flows and lossless compression,'' \emph{Advances in Neural Information Processing Systems}, vol.~32, 2019.

\bibitem{ho2019compression}
J.~Ho, E.~Lohn, and P.~Abbeel, ``Compression with flows via local bits-back coding,'' \emph{Advances in Neural Information Processing Systems}, vol.~32, 2019.

\bibitem{zhang2020ultra}
X.~Zhang and X.~Wu, ``Ultra high fidelity deep image decompression with $\ell_\infty$-constrained compression,'' \emph{IEEE Transactions on Image Processing}, vol.~30, pp. 963--975, 2020.

\bibitem{zhang2021iflow}
S.~Zhang, N.~Kang, T.~Ryder, and Z.~Li, ``iflow: Numerically invertible flows for efficient lossless compression via a uniform coder,'' \emph{Advances in Neural Information Processing Systems}, vol.~34, pp. 5822--5833, 2021.

\bibitem{zhang2021ivpf}
S.~Zhang, C.~Zhang, N.~Kang, and Z.~Li, ``ivpf: Numerical invertible volume preserving flow for efficient lossless compression,'' in \emph{Proceedings of the IEEE/CVF Conference on Computer Vision and Pattern Recognition}, 2021, pp. 620--629.

\bibitem{kang2022pilc}
N.~Kang, S.~Qiu, S.~Zhang, Z.~Li, and S.-T. Xia, ``Pilc: Practical image lossless compression with an end-to-end gpu oriented neural framework,'' in \emph{Proceedings of the IEEE/CVF Conference on Computer Vision and Pattern Recognition}, 2022, pp. 3739--3748.

\bibitem{lc-fdnet}
H.~Rhee, Y.~I. Jang, S.~Kim, and N.~I. Cho, ``Lc-fdnet: Learned lossless image compression with frequency decomposition network,'' in \emph{Proceedings of the IEEE/CVF Conference on Computer Vision and Pattern Recognition}, 2022, pp. 6033--6042.

\bibitem{zhang2024BPS}
Z.~Zhang, H.~Wang, Z.~Chen, and S.~Liu, ``Learned lossless image compression based on bit plane slicing,'' in \emph{Proceedings of the IEEE/CVF Conference on Computer Vision and Pattern Recognition}, 2024, pp. 27\,579--27\,588.

\bibitem{van2016pixel}
A.~Van Den~Oord, N.~Kalchbrenner, and K.~Kavukcuoglu, ``Pixel recurrent neural networks,'' in \emph{International conference on machine learning}.\hskip 1em plus 0.5em minus 0.4em\relax PMLR, 2016, pp. 1747--1756.

\bibitem{salimans2017pixelcnn++}
T.~Salimans, A.~Karpathy, X.~Chen, and D.~P. Kingma, ``Pixelcnn++: Improving the pixelcnn with discretized logistic mixture likelihood and other modifications,'' \emph{arXiv preprint arXiv:1701.05517}, 2017.

\bibitem{kingma2013auto}
D.~P. Kingma and M.~Welling, ``Auto-encoding variational bayes,'' \emph{arXiv preprint arXiv:1312.6114}, 2013.

\bibitem{kobyzev2020normalizing}
I.~Kobyzev, S.~J. Prince, and M.~A. Brubaker, ``Normalizing flows: An introduction and review of current methods,'' \emph{IEEE transactions on pattern analysis and machine intelligence}, vol.~43, no.~11, pp. 3964--3979, 2020.

\bibitem{rippel2017}
O.~Rippel and L.~Bourdev, ``Real-time adaptive image compression,'' \emph{arXiv preprint arXiv:1705.05823}, 2017.

\bibitem{agustsson2019}
E.~Agustsson, M.~Tschannen, F.~Mentzer, R.~Timofte, and L.~V. Gool, ``Generative adversarial networks for extreme learned image compression,'' in \emph{Proceedings of the IEEE International Conference on Computer Vision}, 2019, pp. 221--231.

\bibitem{johnston2018}
N.~Johnston, D.~Vincent, D.~Minnen, M.~Covell, S.~Singh, T.~Chinen, S.~Jin~Hwang, J.~Shor, and G.~Toderici, ``Improved lossy image compression with priming and spatially adaptive bit rates for recurrent networks,'' in \emph{Proceedings of the IEEE Conference on Computer Vision and Pattern Recognition}, 2018, pp. 4385--4393.

\bibitem{li2018}
M.~Li, W.~Zuo, S.~Gu, D.~Zhao, and D.~Zhang, ``Learning convolutional networks for content-weighted image compression,'' in \emph{Proceedings of the IEEE Conference on Computer Vision and Pattern Recognition}, 2018, pp. 3214--3223.

\bibitem{shin2022expanded}
C.~Shin, H.~Lee, H.~Son, S.~Lee, D.~Lee, and S.~Lee, ``Expanded adaptive scaling normalization for end to end image compression,'' in \emph{European Conference on Computer Vision}.\hskip 1em plus 0.5em minus 0.4em\relax Springer, 2022, pp. 390--405.

\bibitem{pan2022content}
G.~Pan, G.~Lu, Z.~Hu, and D.~Xu, ``Content adaptive latents and decoder for neural image compression,'' in \emph{European Conference on Computer Vision}.\hskip 1em plus 0.5em minus 0.4em\relax Springer, 2022, pp. 556--573.

\bibitem{li2022content}
M.~Li, S.~Gao, Y.~Feng, Y.~Shi, and J.~Wang, ``Content-oriented learned image compression,'' \emph{arXiv preprint arXiv:2207.14168}, 2022.

\bibitem{lin2020spatial}
C.~Lin, J.~Yao, F.~Chen, and L.~Wang, ``A spatial rnn codec for end-to-end image compression,'' in \emph{Proceedings of the IEEE/CVF Conference on Computer Vision and Pattern Recognition}, 2020, pp. 13\,269--13\,277.

\bibitem{davd}
X.~Zhang, X.~Wu, X.~Zhai, X.~Ben, and C.~Tu, ``Davd-net: Deep audio-aided video decompression of talking heads,'' in \emph{Proceedings of the IEEE/CVF Conference on Computer Vision and Pattern Recognition}, 2020, pp. 12\,335--12\,344.

\bibitem{minnen2020channel}
D.~Minnen and S.~Singh, ``Channel-wise autoregressive entropy models for learned image compression,'' in \emph{2020 IEEE International Conference on Image Processing (ICIP)}.\hskip 1em plus 0.5em minus 0.4em\relax IEEE, 2020, pp. 3339--3343.

\bibitem{gao2021neural}
G.~Gao, P.~You, R.~Pan, S.~Han, Y.~Zhang, Y.~Dai, and H.~Lee, ``Neural image compression via attentional multi-scale back projection and frequency decomposition,'' in \emph{Proceedings of the IEEE/CVF International Conference on Computer Vision}, 2021, pp. 14\,677--14\,686.

\bibitem{jiang2023mlic}
W.~Jiang, J.~Yang, Y.~Zhai, P.~Ning, F.~Gao, and R.~Wang, ``Mlic: Multi-reference entropy model for learned image compression,'' in \emph{Proceedings of the 31st ACM International Conference on Multimedia}, 2023, pp. 7618--7627.

\bibitem{jiang2023mlicpp}
\BIBentryALTinterwordspacing
W.~Jiang and R.~Wang, ``Mlic++: Linear complexity multi-reference entropy modeling for learned image compression,'' in \emph{ICML 2023 Workshop Neural Compression: From Information Theory to Applications}, 2023. [Online]. Available: \url{https://openreview.net/forum?id=hxIpcSoz2t}
\BIBentrySTDinterwordspacing

\bibitem{wang2023evc}
G.-H. Wang, J.~Li, B.~Li, and Y.~Lu, ``Evc: Towards real-time neural image compression with mask decay,'' \emph{arXiv preprint arXiv:2302.05071}, 2023.

\bibitem{zhang2024olvq}
X.~Zhang and X.~Wu, ``Learning optimal lattice vector quantizers for end-to-end neural image compression,'' \emph{arXiv preprint arXiv:2411.16119}, 2024.

\bibitem{zhu2021transformer}
Y.~Zhu, Y.~Yang, and T.~Cohen, ``Transformer-based transform coding,'' in \emph{International Conference on Learning Representations}, 2021.

\bibitem{lic-tcm}
J.~Liu, H.~Sun, and J.~Katto, ``Learned image compression with mixed transformer-cnn architectures,'' in \emph{Proceedings of the IEEE/CVF Conference on Computer Vision and Pattern Recognition}, 2023, pp. 14\,388--14\,397.

\bibitem{cdc}
R.~Yang and S.~Mandt, ``Lossy image compression with conditional diffusion models,'' \emph{Advances in Neural Information Processing Systems}, vol.~36, 2024.

\bibitem{careil2023towards}
M.~Careil, M.~J. Muckley, J.~Verbeek, and S.~Lathuili{\`e}re, ``Towards image compression with perfect realism at ultra-low bitrates,'' in \emph{The Twelfth International Conference on Learning Representations}, 2023.

\bibitem{ansari1998near}
R.~Ansari, N.~D. Memon, and E.~Ceran, ``Near-lossless image compression techniques,'' \emph{Journal of Electronic Imaging}, vol.~7, no.~3, pp. 486--494, 1998.

\bibitem{zhou2011}
J.~Zhou and X.~Wu, ``$l_2$ restoration of $l_\infty$-decoded images with context modeling,'' in \emph{2011 18th IEEE International Conference on Image Processing}.\hskip 1em plus 0.5em minus 0.4em\relax IEEE, 2011, pp. 1989--1992.

\bibitem{zhou2012}
J.~Zhou, X.~Wu, and L.~Zhang, ``$\ell_2$ restoration of $\ell_\infty$-decoded images via soft-decision estimation,'' \emph{IEEE transactions on image processing}, vol.~21, no.~12, pp. 4797--4807, 2012.

\bibitem{chuah2013}
S.~Chuah, S.~Dumitrescu, and X.~Wu, ``$\ell_2$ optimized predictive image coding with $\ell_\infty$ bound,'' \emph{IEEE transactions on image processing}, vol.~22, no.~12, pp. 5271--5281, 2013.

\bibitem{yuanman2014}
Y.~Li and J.~Zhou, ``Sparsity-driven reconstruction of $\ell_\infty$-decoded images,'' in \emph{2014 IEEE International Conference on Image Processing (ICIP)}.\hskip 1em plus 0.5em minus 0.4em\relax IEEE, 2014, pp. 4612--4616.

\bibitem{florea2016}
R.~Florea, A.~Munteanu, S.-P. Lu, and P.~Schelkens, ``Wavelet-based $l_\infty$ semi-regular mesh coding,'' \emph{IEEE Transactions on Multimedia}, vol.~19, no.~2, pp. 236--250, 2016.

\bibitem{bai2021learning}
Y.~Bai, X.~Liu, W.~Zuo, Y.~Wang, and X.~Ji, ``Learning scalable ly=-constrained near-lossless image compression via joint lossy image and residual compression,'' in \emph{Proceedings of the IEEE/CVF Conference on Computer Vision and Pattern Recognition}, 2021, pp. 11\,946--11\,955.

\bibitem{bai2024deep}
Y.~Bai, X.~Liu, K.~Wang, X.~Ji, X.~Wu, and W.~Gao, ``Deep lossy plus residual coding for lossless and near-lossless image compression,'' \emph{IEEE Transactions on Pattern Analysis and Machine Intelligence}, 2024.

\bibitem{testolina2021towards}
M.~Testolina, E.~Upenik, and T.~Ebrahimi, ``Towards image denoising in the latent space of learning-based compression,'' in \emph{Applications of Digital Image Processing XLIV}, vol. 11842.\hskip 1em plus 0.5em minus 0.4em\relax SPIE, 2021, pp. 412--422.

\bibitem{ranjbar2022joint}
S.~Ranjbar~Alvar, M.~Ulhaq, H.~Choi, and I.~V. Baji{\'c}, ``Joint image compression and denoising via latent-space scalability,'' \emph{Frontiers in Signal Processing}, vol.~2, p. 932873, 2022.

\bibitem{cheng2022optimizing}
K.~L. Cheng, Y.~Xie, and Q.~Chen, ``Optimizing image compression via joint learning with denoising,'' in \emph{European Conference on Computer Vision}.\hskip 1em plus 0.5em minus 0.4em\relax Springer, 2022, pp. 56--73.

\bibitem{huang2023narv}
Y.~Huang, Z.~Duan, and F.~Zhu, ``Narv: An efficient noise-adaptive resnet vae for joint image compression and denoising,'' in \emph{2023 IEEE International Conference on Multimedia and Expo Workshops (ICMEW)}.\hskip 1em plus 0.5em minus 0.4em\relax IEEE, 2023, pp. 188--193.

\bibitem{brummer2023importance}
B.~Brummer and C.~De~Vleeschouwer, ``On the importance of denoising when learning to compress images,'' in \emph{Proceedings of the IEEE/CVF Winter Conference on Applications of Computer Vision}, 2023, pp. 2440--2448.

\bibitem{flicker2k}
B.~Lim, S.~Son, H.~Kim, S.~Nah, and K.~Mu~Lee, ``Enhanced deep residual networks for single image super-resolution,'' in \emph{Proceedings of the IEEE conference on computer vision and pattern recognition workshops}, 2017, pp. 136--144.

\bibitem{kodak24}
R.~Franzen, ``Kodak lossless true color image suite,'' 1999, \url{http://r0k.us/graphics/kodak/}.

\bibitem{div2k}
E.~Agustsson and R.~Timofte, ``Ntire 2017 challenge on single image super-resolution: Dataset and study,'' in \emph{The IEEE Conference on Computer Vision and Pattern Recognition (CVPR) Workshops}, July 2017.

\bibitem{sidd}
A.~Abdelhamed, S.~Lin, and M.~S. Brown, ``A high-quality denoising dataset for smartphone cameras,'' in \emph{Proceedings of the IEEE conference on computer vision and pattern recognition}, 2018, pp. 1692--1700.

\bibitem{adam}
D.~P. Kingma and J.~Ba, ``Adam: A method for stochastic optimization,'' \emph{arXiv preprint arXiv:1412.6980}, 2014.

\bibitem{bm3d}
K.~Dabov, A.~Foi, V.~Katkovnik, and K.~Egiazarian, ``Image denoising by sparse 3-d transform-domain collaborative filtering,'' \emph{IEEE Transactions on image processing}, vol.~16, no.~8, pp. 2080--2095, 2007.

\bibitem{flif}
J.~Sneyers and P.~Wuille, ``Flif: Free lossless image format based on maniac compression,'' in \emph{2016 IEEE international conference on image processing (ICIP)}.\hskip 1em plus 0.5em minus 0.4em\relax IEEE, 2016, pp. 66--70.

\bibitem{restormer}
S.~W. Zamir, A.~Arora, S.~Khan, M.~Hayat, F.~S. Khan, and M.-H. Yang, ``Restormer: Efficient transformer for high-resolution image restoration,'' in \emph{Proceedings of the IEEE/CVF conference on computer vision and pattern recognition}, 2022, pp. 5728--5739.

\bibitem{cgnet}
A.~Ghasemabadi, M.~K. Janjua, M.~Salameh, C.~Zhou, F.~Sun, and D.~Niu, ``Cascadedgaze: Efficiency in global context extraction for image restoration,'' \emph{arXiv preprint arXiv:2401.15235}, 2024.

\end{thebibliography}

\end{document}